\documentclass{aa}  

\usepackage{graphicx}
\usepackage{soul}
\usepackage{txfonts}
\usepackage{float}
\usepackage{times,epsfig,amsmath}
\usepackage{epstopdf}
\usepackage{pdflscape}
\usepackage{adjustbox}
\usepackage{amssymb, amsmath} 
\usepackage[usenames,dvipsnames]{color}
\usepackage{txfonts}
\usepackage{microtype}
\usepackage{url}
\usepackage[bookmarks, colorlinks, breaklinks]{hyperref}  
\hypersetup{linkcolor=blue,citecolor=blue,filecolor=blue,urlcolor=blue} 
\usepackage{booktabs}
\usepackage{subfig}
\usepackage{commath}
\usepackage{natbib}
\usepackage{float}
\usepackage{placeins}
\bibpunct{(}{)}{;}{a}{}{,} 
\usepackage[bookmarks, colorlinks, breaklinks, ]{hyperref}

\begin{document} 
 \title{\textit{In medio stat virtus}: enrichment history in poor galaxy clusters}
 \author{G. Riva
          \inst{1,2}
          \and S. Ghizzardi
          \inst{1}
          \and S. Molendi
          \inst{1}
          \and M. Balboni
          \inst{1,3}
          \and I. Bartalucci
          \inst{1}
          \and S. De Grandi
          \inst{4}
          \and F. Gastaldello
          \inst{1}
          \and\\ L. Lovisari
          \inst{1,5}
          \and M. Rossetti
          \inst{1}
          }

   \institute{INAF - Istituto di Astrofisica Spaziale e Fisica Cosmica di Milano, via A.Corti 12, 20133 Milano, Italy\\
              \email{giacomo.riva@inaf.it}
         \and
             Universit\'a degli Studi di Milano, via G. Celoria 16, 20133 Milano, Italy
         \and 
             Dipartimento di Fisica e Astronomia DIFA – Universit\`a di Bologna, via Gobetti 93/2, 40129 Bologna, Italy 
         \and 
             INAF – Osservatorio Astronomico di Brera, via E. Bianchi 46, 23807, Merate (LC), Italy  
        \and 
             Center for Astrophysics $|$ Harvard $\&$ Smithsonian, 60 Garden Street, Cambridge, MA 02138, USA
             }
\date{Received 24 July 2025 / Accepted 22 November 2025}

\abstract{The enrichment history of galaxy clusters and groups remains far from being fully understood. Recent measurements in massive clusters have revealed remarkably flat iron abundance profiles out to the outskirts, suggesting that similar enrichment processes have occurred for all systems. In contrast, the situation for galaxy groups is less clear: their abundance profiles sometimes appear to decline with radius, challenging our understanding of the physical processes at these scales. In this paper, we present a pilot study aimed at accurately measuring the iron abundance profiles of MKW3s, A2589, and Hydra A, three poor clusters with total masses of $M_{500} \simeq 2.0-2.5 \times 10^{14}$ M$_\odot$, intermediate between the scales of galaxy groups and massive clusters. For these systems, we have also obtained nearly complete azimuthal coverage of their outer regions with XMM-\textit{Newton}, allowing for a detailed characterisation of their chemical properties at large radii. We show that, in the outskirts, abundance measurements are more likely to be limited by systematic uncertainties than by statistical errors. In particular, inaccurate modelling of the soft X-ray background can significantly bias metallicity estimates in regions where the cluster emission is faint. However, once these systematics are properly accounted for, the abundance profiles of all three clusters are found to be consistent with being flat, at a level of $Z \sim 0.3$ Z$_{\odot}$, in agreement with values observed in massive clusters. Using available stellar mass estimates for the three systems, we also computed their iron yields, thereby beginning to probe a mass range that remains largely unexplored. We find $\mathcal{Y}_{\mathrm{Fe,500}} = 2.68 \pm 0.34$, $2.54 \pm 0.64$, and $7.51 \pm 1.47$ Z$_{\odot}$ for MKW3s, A2589, and Hydra A, respectively, values that span the transition regime between galaxy groups and massive clusters. Future observations of systems with temperatures in the $2-4$ keV range will be essential to further populate this intermediate-mass regime and to draw firmer conclusions about the chemical enrichment history of galaxy systems across the full mass scale.}

\maketitle

\section{Introduction}
Galaxy clusters are dark-matter-dominated astrophysical objects with total masses of $\sim 10^{14}-10^{15}$ M$_{\odot}$. They reside at the nodes of the cosmic web and grow by accreting matter as smaller groups fall into their deep gravitational potential wells. Most of the baryons in clusters are located in the so-called intracluster medium (ICM), the diffuse and hot gas at temperatures $T\simeq1-10$ keV that fills the space between cluster galaxies. The ICM is composed primarily of ionised hydrogen and helium, with traces of heavier elements, commonly referred to as metals. Metals are produced through thermonuclear processes occurring in stars at different stages of their evolution \citep[see][for a review]{cameron57,burbidge57,nomoto13}, especially in their final phases, and are subsequently ejected into the ICM up to scales of several Mpc via mechanisms such as stellar winds, AGN feedback, jets, and ram-pressure stripping \citep{schindler08}. While the nucleosynthesis of heavy elements is reasonably well understood, the overall enrichment history of galaxy clusters and groups remains far from clear. In this context, measuring reliable abundance profiles out to their outskirts is crucial, as these regions contain freshly accreted material and preserve the record of cluster formation and evolution processes \citep{walker19}.

In recent years, several studies have investigated the metallicity in the outer regions of massive galaxy clusters, consistently finding nearly constant values of $\sim 0.2-0.4$ Z$_{\odot}$ (solar abundance table from \citealt{asplund09}) beyond the core. Some of these analyses extended only to intermediate radii (e.g. \citealt{degrandi01}; \citealt{leccardi08b}), while others, in a pioneering way, focused on large radii using the \textit{Suzaku} telescope, in some cases reaching $R_{200}$\footnote{$R_{\Delta}$ is defined as the radius within which the mean total mass density equals $\Delta$ times the critical density of the Universe, $\rho_{\text{c}}(z)=3H_0^2 E(z)^2/8\pi G$, with $E(z)=\sqrt{\Omega_\text{m}(1+z)^3 + \Omega_{\Lambda}}$ for a flat Universe. $M_{\Delta}$ is the total mass enclosed within $R_{\Delta}$.}, along selected azimuthal sectors of the ICM in individual clusters \citep[e.g.][]{werner13,simionescu17} or using archival samples \citep{urban17}. Together, these results provided observational support for the so-called `early enrichment' scenario, which is currently considered as the most plausible explanation for the chemical enrichment of large-scale structures (see e.g. \citealt{mernier18a}; \citealt{biffi18}, for reviews). More recently, \citet{ghizzardi21} presented robust iron-abundance profiles extending to $R_{500}$ for a representative sample of 12 massive clusters ($M_{500} \ge 3.5 \times 10^{14}$ M$_{\odot}$, $T \ge 4$ keV), exploiting the full azimuthal coverage of the ICM ensured by the XMM-\textit{Newton} Cluster Outskirts Project (X-COP; \citealt{eckert17}). Beyond the core, their profiles were found to be remarkably flat out to $\sim R_{500}$, with a small scatter ($\sim 15\%$) around an average value of $\sim 0.37$ Z$_{\odot}$. These results proved robust against possible systematics in the outskirts and are regarded as a solid reference for the abundance level in the outer regions of massive systems, further suggesting that similar enrichment processes have occurred for all systems from $z\sim 2$ to today (e.g. \citealt{ettori15, mcdonald16,mantz20,flores21,molendi24}).

In contrast, the picture is less clear for galaxy groups ($M_{500} \leq 1.5\times 10^{14}$ M$_{\odot}$, and $T \leq 2$ keV). While some studies have reported similarly flat abundance profiles beyond $\sim0.3\,R_{500}$ \citep[e.g.][]{mernier17,lovisari19}, in some cases reaching the extremely challenging limit of $R_{200}$ \citep[e.g.][]{tholken16,sarkar22}, others have found steadily declining profiles, reaching values as low as $\sim0.1-0.2$ Z$_\odot$ at large radii \citep[e.g.][]{rasmussen07,rasmussen09,sun12}. These discrepancies likely reflect the complex baryonic physics at play in low-mass systems. Due to their relatively shallow gravitational potential wells, galaxy groups are more strongly affected by non-gravitational processes than massive clusters. In particular, AGN feedback plays a major role in mixing and expelling metals to large distances, thereby hampering the compilation of a full inventory of the metal content in these systems \citep[see e.g.][for a review]{gastaldello21}. Consequently, the details of the chemical enrichment history at the low-mass end remain uncertain and are still being debated.

Further challenges arise when considering the effective iron yields, $\mathcal{Y}_{\text{Fe}}$ \citep{renzini14}. We define the iron yield as the ratio between the total iron mass within a given radius, including both the iron dispersed in the ICM and that still locked in stars, and the stellar mass that produced it. As such, it quantifies the efficiency of iron production by the stellar populations in groups and clusters. \citet{ghizzardi21,ghizzardi21-corrigendum} measured the iron yields within $R_{500}$ for seven clusters in the X-COP sample and found a discrepancy of a factor of $3-7$ with respect to theoretical expectations based on supernova explosions \citep{renzini14,freundlich21}, suggesting that the stars within these clusters may not be sufficient or efficient enough to account for all the observed iron. The observed discrepancies at the most massive cluster scale have also been reinforced by recent work based on simulated clusters \citep{biffi25}. Conversely, good agreement between observational results and theoretical expectations was derived for a sample of galaxy groups \citep{renzini14}, further highlighting the distinction between groups and massive clusters and posing the basis for the so-called `Fe Conundrum'. However, we should note that the low gas fraction observed in these low-mass systems directly affects the results obtained at this scale \citep[see e.g.][]{eckert21,gastaldello21}.

Useful information on the metal enrichment of hot halos can be provided by systems with masses $M_{500} \simeq 1.5-3.5 \times 10^{14}$ M$_{\odot}$, and typical temperatures $T \simeq 2-4$ keV, i.e. the missing link between the group and the massive cluster scale. A comprehensive study of the radial abundance profiles extending to the outermost regions of these poor clusters is crucial to determine whether they are consistent with measurements performed in more massive systems, or whether significant discrepancies begin to emerge. For example, close agreement with the results for massive systems would suggest that similar enrichment processes have occurred at all scales. Conversely, results similar to those occasionally observed in groups would highlight the need for more comprehensive models capable of explaining these differences. In addition, the study of these intermediate-mass systems provides an opportunity to further investigate the `Fe Conundrum', since the $M_{500} \simeq 1.5-3.5 \times 10^{14}$ M$_{\odot}$ mass range remains largely unexplored in terms of iron-yield measurements. The inclusion of new observational results in this range is therefore crucial to either validate or revise currently proposed models on this topic.

However, deriving reliable abundance measurements in such poor clusters is particularly challenging. At temperatures $T \simeq 2-4$ keV, the Fe K$\alpha$ transitions at $\sim 6.7$ keV are no longer the primary diagnostic for deriving abundances, as the Fe L-shell lines at $\sim 1$ keV become brighter. The large number of Fe emission lines peaking in the $0.9-1.3$ keV energy range \citep[see][]{hwang97,mazzotta98,gu19,gu20,heuer21}, the limited spectral resolution of the currently available CCD cameras, and our current knowledge of the atomic processes, all make the modelling of the L-shell challenging, thus questioning the reliability of abundance measurements based on it alone.  In the central regions of these systems, where the source emission exceeds the background contribution, the intensities of the Fe L-shell and K$\alpha$ lines are comparable, and the former is not a source of any systematic errors \citep{riva22}. Conversely, in the external cluster regions, the increasing contribution from the XMM-\textit{Newton} instrumental background makes the Fe L-shell the only viable option for abundance measurements. Under these conditions, these measurements are further complicated by the contamination from the soft X-ray background (XRB), which is commonly divided into the cosmic X-ray background \citep[CXB;][]{giacconi01} and the local galactic foregrounds \citep[i.e. Local Hot Bubble, LHB; Galactic Halo, GH; North Polar Spur, NPS; see e.g.][]{snowden97,kuntz00}. Inaccurate modelling of these components can also potentially bias the results obtained at this mass scale.

In this paper we present a pilot study of three poor clusters with masses of $M_{500} \simeq 2.0-2.5 \times 10^{14}$ M$_{\odot}$, i.e. MKW3s, A2589 and Hydra A, for which we have also obtained proprietary XMM-\textit{Newton} offset observations to ensure azimuthal coverage of their outermost regions. Our aim is to derive robust abundance profiles of these systems out to their outskirts, with a particular focus on assessing the systematic uncertainties associated with the characterisation of the XRB. The results presented in this work provide a solid baseline for a future, more extensive project focusing on the metal enrichment of intermediate-mass systems. 

The paper is structured as follows: in Sect. \ref{sample}, we present the sample used for this study; in Sect. \ref{data_pro}, we outline the key aspects of our analysis procedure, including our novel approach to estimating XRB contamination; we present the results of the work in Sect. \ref{results}, and discuss our findings in Sect. \ref{discussion}; finally, we present our conclusions in Sect. \ref{conclusion}. Throughout the paper, we assume a $\Lambda$ cold dark matter (CDM) cosmology with $H_0 = 70$ km s$^{-1}$ Mpc$^{-1}$, $\Omega_{\text{m}} = 0.3$, $\Omega_{\Lambda} = 0.7$, and $E(z) =\sqrt{\Omega_{\text{m}}(1 + z)^3 + \Omega_{\Lambda}}$ for the evolution of the Hubble parameter. The solar abundance table is set to \citet{asplund09}. Unless stated otherwise, all the quoted errors are at the $1\sigma$ confidence level. At the redshifts of the clusters, $1'$ corresponds to $\sim 53$, $49$, and $63$ kpc for MKW3s, A2589, and Hydra A, respectively.


\begin{table*}
\caption{\footnotesize Cluster properties and information on the used XMM-\textit{Newton} observations.}
\centering 
\begin{tabular}{c|cccccc|cccc}
\toprule   
\toprule
Name & RA & DEC & $N_{\mathrm{H}}$ & $z$ & $M_{500}$ & $R_{500}$ & Obs. IDs & Type & $t_{\mathrm{exp}}$ & XRB regions \\
& (deg) & (deg) & ($10^{20}$ cm$^{-2}$) & &($10^{14}$ M$_{\odot}$) & (arcmin) & & & (ks) & \\
\midrule  
 & & & & & & & $0723801501$ & Central & $93$ & -- \\
 & & & & & & & $0109930101$ & Central & $30$ & -- \\
MKW3s & $230.464$& $7.706$ & $2.80$ & $0.045$& $2.41$ & $17.6$ & $0904610201$ & Offset & $27$ & bkg1, bkg2 \\
 & & & & & & & $0904610101$ & Offset & $28$ & bkg3, bkg4 \\
 & & & & & & & $0904610401$ & Offset & $30$ & bkg5, bkg6 \\
 & & & & & & & $0904610301$ & Offset & $31$ & bkg7, bkg8 \\
 \midrule
 & & & & & & & $0204180101$ & Central & $20$ & -- \\
A2589 & $350.987$ & $16.766$ & $2.96$ & $0.041$ & $2.59$ & $19.7$ & $0924120301$ & Offset & $27$ & bkg1, bkg2 \\
 & & & & & & & $0924120201$ & Offset & $27$ & bkg3, bkg4 \\
\midrule  
 & & & & & & & $0843890101$ & Central & $77$ & -- \\
 & & & & & & & $0843890201$ & Central & $90$ & -- \\
& & & & & & & $0843890301$ & Central & $87$ & -- \\
Hydra A & $139.524$ & $-12.096$ & $4.68$ & $0.054$ & $1.98$ & $13.9$ & $0504260101$ & Central & $54$ & -- \\
 & & & & & & & $0694440401$ & Offset & $25$ & bkg1, bkg2 \\
 & & & & & & & $0694440701$ & Offset & $14$ & bkg3, bkg4 \\
 & & & & & & & $0694440801$ & Offset & $17$ & bkg5, bkg6 \\
\bottomrule                                  
\end{tabular}
\tablefoot{In the table we list the names of the clusters, the coordinates of their X-ray peaks, the Galactic absorption $N_{\mathrm{H}}$ along the direction of each cluster \citep{HI4PI16}, the cluster redshifts \citep{piffaretti11}, and also the $M_{500}$ and $R_{500}$ values presented in Sect. \ref{sect_mass}. We report the ID of the observations used for this study, and whether they are central or offset pointings of the clusters. We also provide the average exposure time $t_{\mathrm{exp}}$ of each observation after soft proton cleaning, weighting the exposure of the three cameras as $t_{\mathrm{exp}} = 0.25 \times t_{\mathrm{exp,MOS1}} + 0.25 \times t_{\mathrm{exp,MOS2}} + 0.5 \times t_{\mathrm{exp,pn}}$. Finally, we also list the names of the regions extracted from each offset observation to estimate the XRB parameters (see Fig. \ref{clusters}). The south and west offset observations of A2589 are discarded as heavily contaminated by soft proton flares, while the south-east offset pointing of Hydra A is discarded due to the diffuse thermal emission from LEDA 87445 (see Sect. \ref{data_red}).}
\label{table:1}
\end{table*}

\section{The sample}
\label{sample}
We have selected a pilot sample of three nearby, bright galaxy clusters, namely MKW3s, A2589 and Hydra A, with masses in the intermediate range of interest. The combination of their masses with their low redshifts allows spatially resolved spectroscopic analysis out to their outermost regions, using one central and four offset XMM-\textit{Newton} pointings. In particular:
\begin{itemize}
    \item MKW3s is a poor, cool-core cluster located at $z = 0.044$, with a mass of $M_{500} \simeq 1.5-2.5 \times 10^{14}$ M$_\odot$ \citep{piffaretti11,zhang11,kravtsov18}. Deep central observations from XMM-\textit{Newton} (a total exposure of $\sim 175$ ks) are publicly available for this cluster. After AO-21, we were awarded four additional offset pointings (PI: S. Ghizzardi), using the same strategy as used in the X-COP project \citep[see][]{eckert17}. These new observations, with a total exposure of $\sim 160$ ks, provide full azimuthal coverage of the cluster outskirts, allowing accurate modelling of the XRB components and thus reliable metallicity measurements out to $R \gtrsim 0.8 R_{500}$;
    \item A2589 is another nearby and poor cluster, located at $z = 0.041$ and with a reported mass of $M_{500} \simeq 2.0-3.0 \times 10^{14}$ M$_\odot$ \citep{piffaretti11,zhang11}. A central XMM-\textit{Newton} observation of $\sim 50$ ks is available in the XMM-\textit{Newton} Science Archive (XSA), and allows the investigation of the ICM properties in the inner and intermediate regions. In order to study the cluster outskirts in detail and to properly constrain the XRB contribution over the entire cluster extension, we obtained four offset observations also for this cluster ($\sim 60$ ks each; PI: S. Ghizzardi, AO-22), following the same observing strategy as for MKW3s;
    \item Hydra A is a bright cool-core cluster at $z = 0.054$ with an estimated mass of $M_{500} \simeq 1.5-3.5 \times 10^{14}$ M$_\odot$ \citep{piffaretti11,sato12,ettori19}. The XSA contains four deep central observations, each with an unfiltered exposure time exceeding $100$ ks, as well as four additional offset pointings obtained for a pilot programme of the X-COP project \citep{eckert17}. These observations provide a comprehensive coverage of the thermal emission of this cluster out to large radii.
\end{itemize}

\section{Analysis procedures}
\label{data_pro}
For each cluster, we used a combination of archival XMM-\textit{Newton} observations retrieved from the XSA and proprietary data obtained in AO-21 and AO-22 (PI: S. Ghizzardi), as summarised in Table \ref{table:1}. A comprehensive description of our data reduction, imaging, and spectral analysis techniques can be found in \citet{bartalucci+23} and \citet{rossetti24}, which are based on the analysis procedures outlined in \citet{ghirardini19} with the introduction of some novelties. Below we summarise the main steps of our analysis, with particular emphasis on our strategy for estimating the XRB contamination.

\subsection{Data reduction}
\label{data_red}
The three clusters in our sample were observed with the European Photon Imaging Camera (EPIC; \citealt{struder01}; \citealt{turner01}) on board XMM-\textit{Newton}. All datasets were reprocessed using the Extended Science Analysis System \citep[ESAS;][]{snowden08} embedded in SAS version 16.1, keeping the calibration database (CalDB) updated to ensure that we apply the latest calibration files to our data. We refer to \citet{bartalucci+23} for details on the first steps of the data reduction, including calibration, standard pattern cleaning, removal of noisy CCDs in the MOS cameras, and light-curve filtering to remove time intervals affected by soft proton (SP) flares. The resulting exposure times after this procedure are given in Table \ref{table:1} for each observation. During the pre-processing of the data we also computed IN/OUT and IN-OUT indicators, which quantify the residual contamination from soft protons \citep[see e.g.][]{deluca04,leccardi08b,salvetti17,marelli17,marelli21,gastaldello22,rossetti24}. 

We note that the southern (ID: $0924120101$) and western (ID: $0924120401$) offset observations of A2589 were discarded from the analysis due to remarkably high IN/OUT values (IN/OUT $\sim3.4$ and $\sim2.4$, respectively), indicating severe residual SP contamination. In addition, we did not consider any of the several available pointings to the south-east of Hydra A, as LEDA 87445's diffuse thermal emission would prevent an unbiased determination of the cluster properties along this direction \citep[e.g.][]{degrandi16}.

\subsection{Image analysis}
\begin{figure}
            \centering
            \hspace{0.15cm}\includegraphics[width=0.42\textwidth]{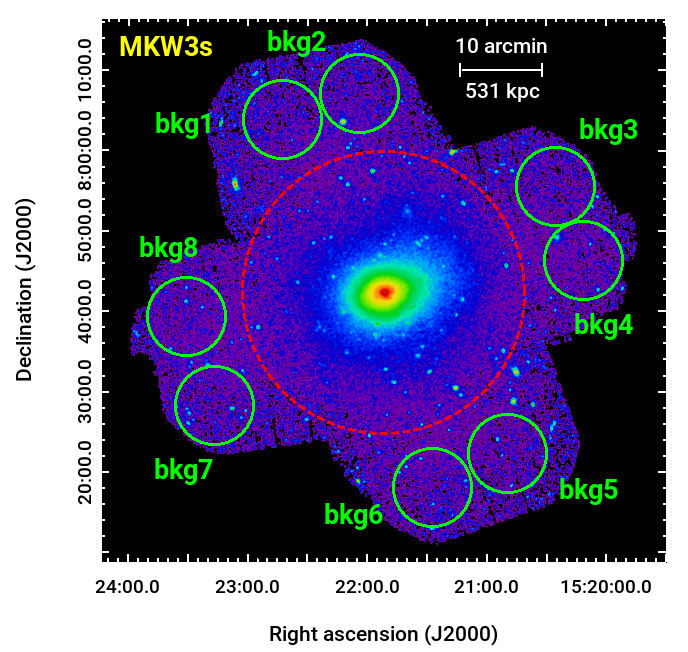}\\
            \hspace{0.1cm}\includegraphics[width=0.42\textwidth]{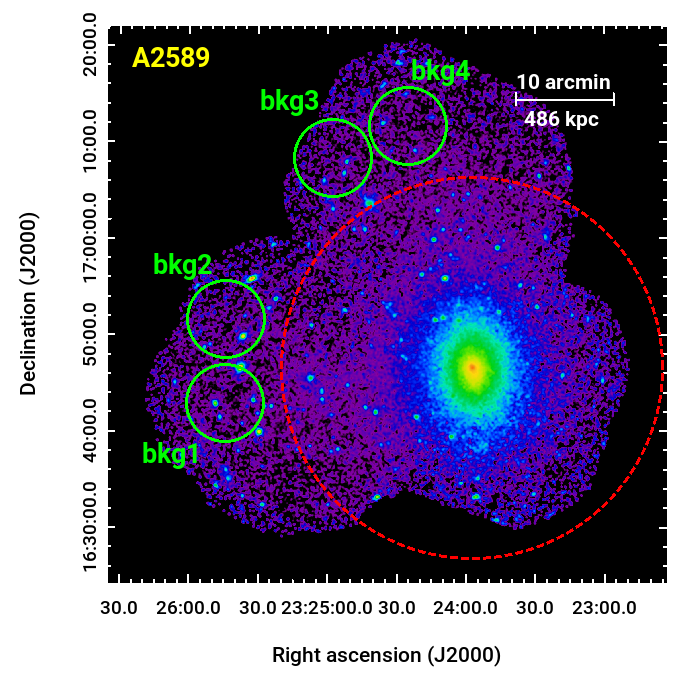} \\
            \hspace{0.15cm}\includegraphics[width=0.42\textwidth]{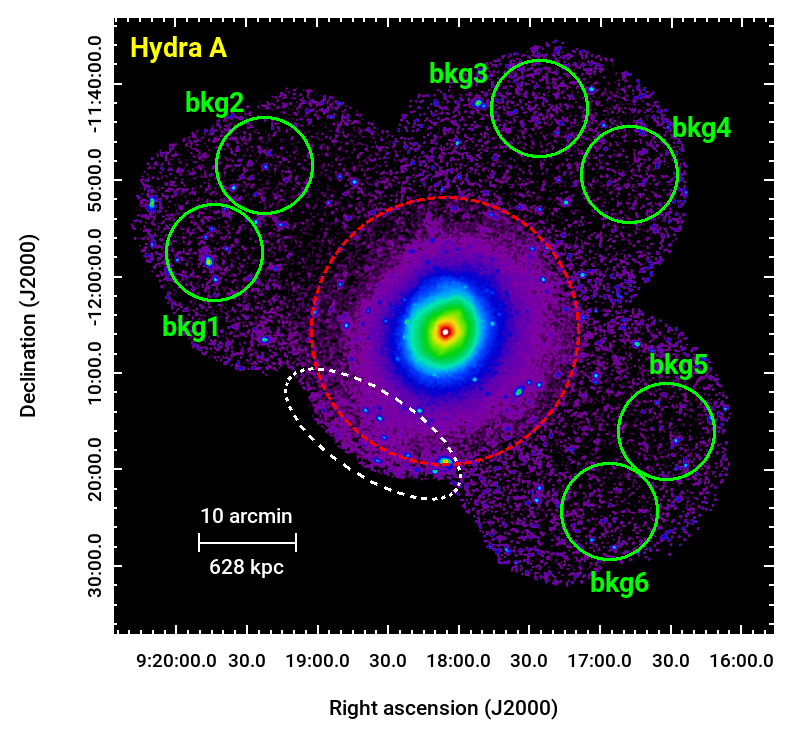}\\
            \caption{\footnotesize XMM-\textit{Newton}/EPIC count-rate images of MKW3s (top), A2589 (centre), and Hydra A (bottom) in the $0.7-1.2$ keV band. The red dashed circles mark $R_{500}$ for each cluster. The green circles are the regions used to estimate the contamination from the X-ray sky background. The white dashed region in the bottom panel has been masked to remove residual thermal emission associated with LEDA 87445. A Gaussian filter smooths the images for visual purposes only.}
            \label{clusters}
\end{figure}

To produce the scientific images of the clusters, we first extracted photon count images from the three XMM-\textit{Newton}/EPIC detectors in the $0.7-1.2$ keV band. We also computed the corresponding exposure maps, which account for vignetting effects, and produced rescaled particle-induced background images using the unexposed corners of the detectors, including contributions from both cosmic-ray particle-induced background and residual soft protons. All the EPIC source, exposure, and background images for each cluster were combined to produce final background-subtracted count rate mosaic images (Fig. \ref{clusters}). 

Point sources within the field of view were identified following the scheme detailed in Sect. 2.2.3 of \citet{ghirardini19}. All detected sources were then visually inspected to check for false detections, such as in the CCD gaps. In the case of Hydra A, in addition to excluding the south-east XMM-\textit{Newton} pointing (see Sect.~\ref{data_red}), we also masked a large region (white in Fig. \ref{clusters}, bottom) in the south-east direction to remove residual thermal emission associated with LEDA 87445.

\subsection{Spectral analysis}
\label{spec_analysis}
Following \citet{eckert14}, we have adopted a full spectral modelling approach for all components, including the XMM-\textit{Newton} particle-induced background, the soft X-ray background, and the source emission. Briefly, our procedure for the spectral analysis of the three clusters involves two sequential steps, which we outline in more detail below. First, we identify regions distant enough from the cluster thermal emission to measure XRB spectral parameters, which we assume are representative of the contamination across the entire cluster extension. Second, we apply the best-fit XRB model to spectra extracted from cluster regions and use this to recover radial information about the plasma properties of the three clusters, including their temperatures and metal abundances.

\subsubsection{Regions and adopted strategy}
As discussed in more detail in Sects. \ref{results} and \ref{variability}, a key challenge in the analysis of systems with $T \simeq 2-4$ keV, and even lower, lies in the robustness of the modelling of the XRB components. Indeed, even small inaccuracies in these background parameters can significantly bias the reconstructed properties of a cluster. 

To further investigate these potential systematics, we adopted a novel approach and modelled the XRB contamination from multiple regions placed azimuthally around each cluster. Specifically, we extracted spectra from $8$, $4$, and $6$ circular regions around MKW3s, A2589, and Hydra A (i.e. two regions per offset observation; see Fig. \ref{clusters}), located beyond $r = 20'$, $22'$, and $20'$, respectively. To identify these radial ranges, we used \textit{pyproffit} \citep{eckert20} to extract surface brightness profiles from particle background–subtracted images of the three clusters in the $0.7-1.2$ keV band, which maximises the source-to-background (S/B) ratio \citep{ettori10}. Since these profiles include contributions from both the source and the XRB, we identified the radii where they flatten, indicating that the source emission becomes negligible compared to the local XRB beyond these thresholds\footnote{Even in the favourable $0.7-1.2$ keV energy band, the estimated source emission in the selected XRB regions is less than $\sim 1\%$, $2\%$, and $5\%$ for MKW3s, A2589, and Hydra, respectively.}. The circular regions have radii of $5'$ for MKW3s and Hydra A, and are slightly smaller (i.e., $r = 4'$) for A2589. As with the use of a single large annulus covering the outskirts of the XMM-\textit{Newton} mosaic, the adoption of multiple regions allows us to derive XRB parameters that are averaged\footnote{In the following, we will indicate the XRB parameters obtained from a joint fit of the circular regions as `azimuthally averaged'. We specify that the joint fitting procedure is technically different from a simple average of the parameters derived from individual fits, although they provide similar results.} over the full azimuthal extent, through a joint fit of all extracted spectra (Sect. \ref{azimuthal_aver}). However, the adopted technique also enables us to go a step further. By applying the different XRB models individually, we can investigate how the choice of a specific region affects the temperature and abundance measurements, particularly in their cluster outskirts, where the source intensities are low (Sect. \ref{individ_meas}).

Cluster spectra were extracted from concentric annuli centred on the X-ray peak of each system (Table \ref{table:1}), following this general scheme: i) annuli $0.5'$ wide within $3'$; ii) $1'$ wide for $3' < r < 6'$; iii) $2'$ wide between $6'$ and $12'$; iv) one additional bin between $12'$ and $15'$ for all systems; and v) a final bin out to $18'$ for A2589. Based on the reconstructed cluster masses (Sect. \ref{sect_mass}), the outermost bins reach $\sim 0.85$, $\sim 0.91$, and $\sim 1.08~R_{500}$ for MKW3s, A2589, and Hydra A, respectively. We explored the possibility of extending the measurements further into the outskirts for MKW3s; however, this cluster lies on the NPS \citep{snowden97,merloni24}, a region of the sky affected by strong XRB contamination, which makes reliable measurements at large radii particularly difficult. Similarly, in the outermost bin of A2589 and Hydra A, the very low source signal (S/B $< 0.1$; across the entire energy range) in these regions, makes the abundance measurements based on XRB modelling from individual background regions difficult to constrain in some cases. For these two clusters, we therefore considered only the results obtained from azimuthally averaged XRB values. Nevertheless, as shown in Sect.~\ref{results}, the available radial coverage is sufficient to derive meaningful conclusions for all clusters.

\subsubsection{Spectral extraction and fitting procedures}
We extracted spectra, redistribution matrix files, and ancillary response files for each region using the ESAS tools \texttt{mos-spectra} and \texttt{pn-spectra}, which also select the most appropriate filter-wheel-closed data based on magnitude and hardness ratio \citep[see][]{kuntz08}. The corresponding particle-induced background spectra were generated using \texttt{mos-back} and \texttt{pn-back}, and used to construct the particle background model, following the procedure described in \citet{rossetti24}.

For each region, we used XSPEC v.12 \citep{arnaud96} to perform a joint fit of the spectra extracted from the three EPIC detectors and from multiple observations covering the same area. We used a standard maximum likelihood approach, using the Cash statistics \citep{cash79}. Errors are then calculated according to variations in Cash statistics. We fitted the XRB spectra in the $0.5-12$ keV and $0.5-14$ keV ranges for MOS and pn detectors, respectively. While including the soft $0.5-0.7$ keV energy range is necessary to constrain the XRB spectral properties, particularly the normalisation of the LHB, for the source fitting we increased the minimum energy to $E_{\mathrm{min}} = 0.7$ keV. This choice helps reduce the systematics associated with inaccurate XRB modelling in the final cluster measurements (Sect. \ref{variability}). We also excluded from the fitting procedure the energy ranges where prominent fluorescence lines are present, i.e. $1.2 - 1.9$ keV for MOS detectors, $1.2 - 1.7$ keV and $7 - 9.2$ keV for pn. 

\begin{figure}
            \centering
            \includegraphics[width = 0.5\textwidth]{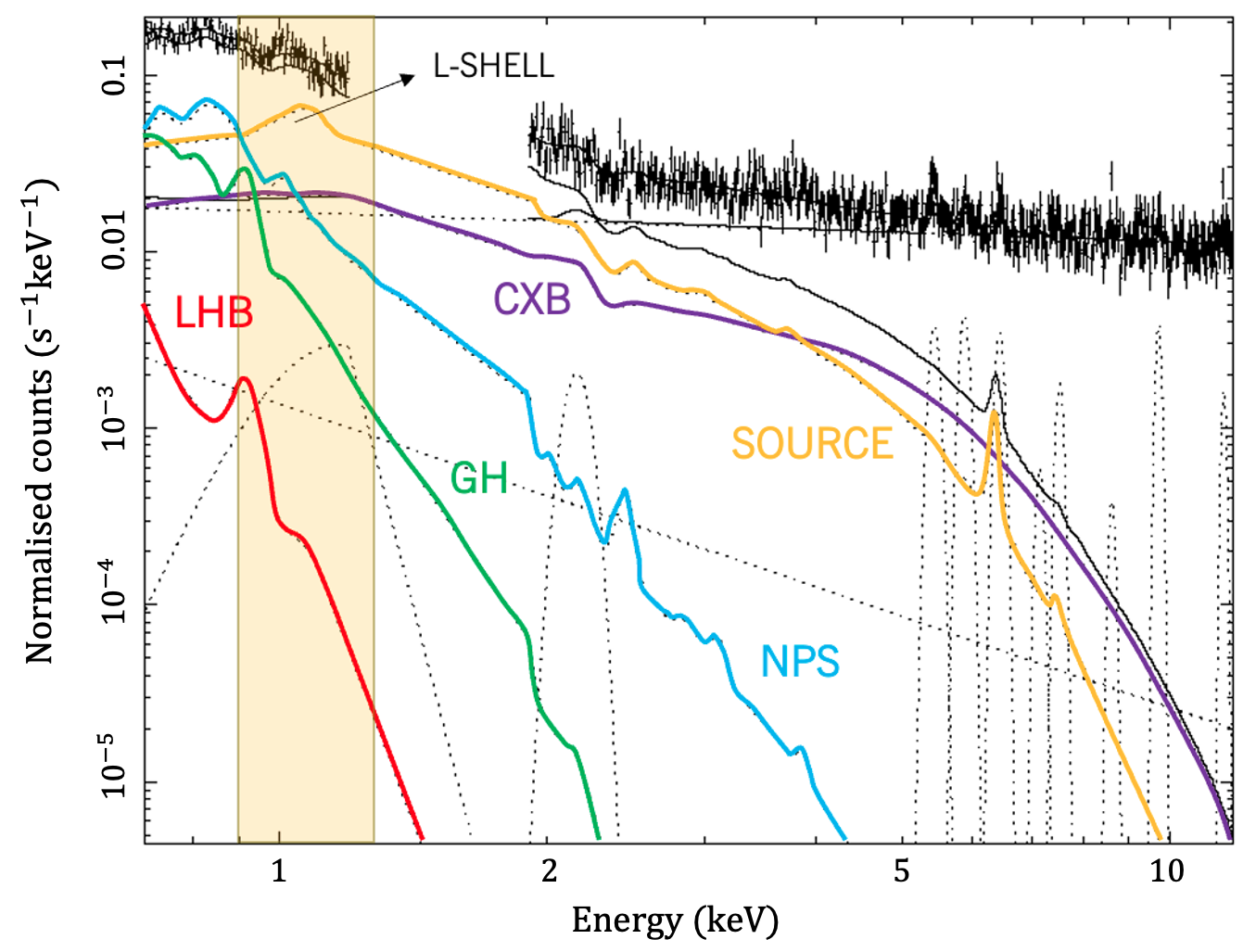}
            \caption{\footnotesize Relative contribution between source emission and background components, in the outskirts of MKW3s. Black data are the XMM-\textit{Newton}/MOS 2 spectrum, extracted from the ring with $r = 10' - 12'$ (i.e. $0.57-0.68~R_{500}$). Yellow is used to highlight the best-fitting model for the source emission; the Fe L-shell blend is visible at $\sim 1$ keV. Red, green, light blue, and purple are used to mark the LHB, the GH, the NPS, and the CXB, estimated from the region named bkg1. Dotted lines are the model adopted to describe the particle-induced background, which includes power laws and Gaussian lines. The solid black lines show the resulting model for the X-ray sky components (i.e. the source plus the XRB) and the particle-induced background.}
            \label{did_pic}
\end{figure}

We adopted a combination of different models accounting for the source emission, the soft X-ray background, and the XM-\textit{Newton} instrumental background. These are detailed below:
\begin{itemize}
    \item we modelled the cluster thermal emission using \texttt{apec} \citep[Astrophysical Plasma Emission Code;][]{smith01}, and accounted for Galactic absorption with  \texttt{phabs}. Cluster temperature, metal abundance, and normalisation were left free to vary, while the redshift and absorption were fixed to the values listed in Table~\ref{table:1};
    \item to model the XRB contamination, we used an unabsorbed \texttt{apec} component for the LHB, with temperature fixed at $0.11$ keV, and an absorbed thermal model (\texttt{phabs*apec}) with temperature allowed to vary between $0.1$ and $0.6$ keV for the GH \citep{mccammon02}. Given the projection of MKW3s onto the NPS, we also included an additional thermal component with free temperature for this cluster \citep[see e.g.][]{markevitch03,vikhlinin05,miller08}. We modelled the CXB with an absorbed power law, using a fixed photon index of $1.46$ \citep{moretti09}. For all the XRB models, the normalisations were left free to vary, while we fixed the redshifts to zero, used solar abundances and the same absorption as for the cluster emissions;
    \item finally, we also applied a tailored model for the particle-induced background for each detector, consisting of two power laws and a set of Gaussian fluorescence lines. An additional power law was included to model residual soft proton contamination, based on the latest IN-OUT calibration provided by \citet{rossetti24}.
\end{itemize}  
As already pointed out, we follow a two-step procedure. First, we fit the XRB model to the background regions. Then, we apply the resulting background model, keeping its parameters fixed to the best fit values obtained in the previous step, in combination with the source model to fit the cluster spectra. In both cases, we also include the best-fit model for the XMM-\textit{Newton} instrumental background specific to each region. The relative contributions of the various source and background components are illustrated in Fig. \ref{did_pic}, which shows the best fit in one of the outer annuli of MKW3s as an example.

Finally, we also verified that there is no significant contamination from solar wind charge exchange (SWCX; see \citealt{kuntz19} for a review) in the offset observations of the three clusters, where the presence of this astrophysical nuisance could significantly bias our reconstruction of the best-fit XRB models. For both A2589 and Hydra A, the average proton flux measured by the \textit{Advanced Composition Explorer}/SWEPAM\footnote{Data available at the following link:\\ \href{https://izw1.caltech.edu/ACE/ASC/level2/swepam_l2desc.html}{https://izw1.caltech.edu/ACE/ASC/level2/swepam\_l2desc.html}.} instrument during the observations was below $\sim 4 \times 10^8$ protons cm$^{-2}$ s$^{-1}$, which is typical of the quiescent Sun \citep{snowden04}. In the case of MKW3s, one observation showed an average proton flux near this threshold, while the remaining observations were safely below it. However, given the presence of additional contamination from the NPS, any possible SWCX contribution is a subdominant component at low energies for this cluster.

\subsection{New mass measurements}
\label{sect_mass}
Measurements of $M_{500}$ (and consequently $R_{500}$) for MKW3s, A2589, and Hydra A are available in the literature \citep[e.g.][]{piffaretti11,zhang11,kravtsov18,sato12,ettori19}. However, to ensure consistency with the density and temperature profiles derived in this work, and to provide homogeneous estimates across our sample, we recalculated the mass values for all three clusters. 

We used the \textit{hydromass} code \citep{eckert22a} to derive new estimates of the total masses for MKW3s, A2589, and Hydra A under the assumption of hydrostatic equilibrium, which we will adopt in the following sections to rescale the radial profiles accordingly. For this purpose, we assumed an NFW model for the total mass \citep{NFW96}, and also used the spectroscopic profiles obtained with the azimuthally averaged XRB parameters (see Sect. \ref{azimuthal_aver}). The resulting total masses are $M_{500} = 2.41 \pm 0.04$, $2.59 \pm 0.09$, and $1.98 \pm 0.01 \times 10^{14}$ M$_{\odot}$ for MKW3s, A2589, and Hydra A, respectively, consistent with the available measurements for these systems (Sect. \ref{sample}). From the measured masses, we derived $R_{500} \simeq 933$, $956$, and $871$ kpc, respectively, with the corresponding values in arcminutes listed in Table \ref{table:1}. 


\section{Spectral results}
\label{results}
In this section, we present the spectral results obtained for the three clusters. We begin by discussing the temperature and abundance profiles derived using an azimuthally averaged characterisation of the XRB contribution. We then compare these results with those obtained by modelling the XRB from individual background regions.

\subsection{Azimuthally averaged measurements}
\label{azimuthal_aver}

\begin{figure*}
            \centering
            \includegraphics[width = \textwidth]{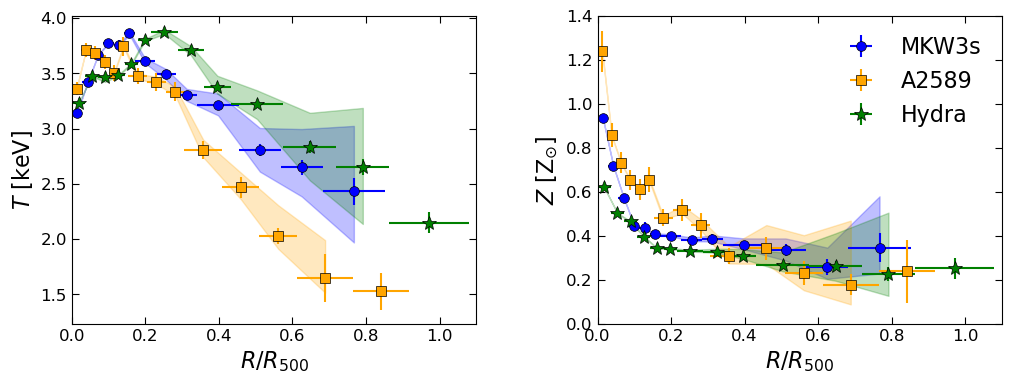}
            \caption{Temperature (left) and iron abundance (right) profiles for MKW3s, A2589 and Hydra A. Blue dots, orange squares, and green stars represent measurements obtained with azimuthally averaged XRB parameters. The effect of modelling the XRB contamination from individual regions is shown by the shaded areas, which capture the maximum dispersion of the measured profiles shown in Fig. \ref{individual_prof}. As mentioned in the text, the measurements derived from individual XRB modelling for both A2589 and Hydra A do not extend to the outermost radial bins due to the very low source signal which makes them difficult to constrain.}
            \label{spec_prof}
\end{figure*}

We first used all available spectra from the offset observations of MKW3s, A2589 and Hydra A to perform a joint fit of the XRB parameters. These measurements are shown as the grey bands in Figs. \ref{XRB_mkw3s}, \ref{XRB_a2589}, and \ref{XRB_hydra}, and provide the best representation of the XRB contamination over the cluster extensions, allowing reliable cluster measurements into their outer regions. The temperature and iron abundance profiles obtained from this specific background characterisation are shown in Fig. \ref{spec_prof} (left and right, respectively) as the blue dots, orange squares and green stars, each rescaled to the $R_{500}$ values derived in the previous section. These abundance measurements are also given in Table \ref{table:2} as $Z_{\mathrm{joint}}$.

\begin{table*}
\caption{\footnotesize Iron abundance measurements for MKW3s, A2589, and Hydra A.}
\centering 
\begin{tabular}{c|ccc|ccc|ccc}
\toprule   
\toprule
&& MKW3s &&& A2589 &&& Hydra A & \\
\midrule
Annulus & $Z_{\mathrm{joint}}$ & $Z_\mathrm{low}$ & $Z_{\mathrm{high}}$ & $Z_{\mathrm{joint}}$ & $Z_\mathrm{low}$ & $Z_{\mathrm{high}}$ & $Z_{\mathrm{joint}}$ & $Z_\mathrm{low}$ & $Z_{\mathrm{high}}$ \\
(arcmin)& (Z$_{\odot}$) & (Z$_{\odot}$) & (Z$_{\odot}$) & (Z$_{\odot}$) & (Z$_{\odot}$) & (Z$_{\odot}$) & (Z$_{\odot}$) & (Z$_{\odot}$) & (Z$_{\odot}$) \\
\midrule  
$0.0-0.5$ & $0.94 \pm 0.02$ & $0.94$ & $0.94$ & $1.24 \pm 0.09$ & $1.24$ & $1.24$ & $0.62 \pm 0.01$ & $0.62$ & $0.62$ \\
$0.5-1.0$ & $0.72 \pm 0.01$ & $0.72$ & $0.72$ & $0.86 \pm 0.05$ & $0.85$ & $0.86$ & $0.50 \pm 0.01$ & $0.50$ & $0.50$ \\
$1.0-1.5$ & $0.57 \pm 0.01$ & $0.57$ & $0.57$ & $0.73 \pm 0.05$ & $0.73$ & $0.73$ & $0.46 \pm 0.01$ & $0.46$ & $0.47$ \\
$1.5-2.0$ & $0.45 \pm 0.01$ & $0.45$ & $0.45$ & $0.65 \pm 0.04$ & $0.65$ & $0.66$ & $0.40 \pm 0.01$ & $0.40$ & $0.40$ \\
$2.0-2.5$ & $0.43 \pm 0.03$ & $0.43$ & $0.44$ & $0.61 \pm 0.05$ & $0.61$ & $0.61$ & $0.34 \pm 0.01$ & $0.34$ & $0.34$ \\
$2.5-3.0$ & $0.41 \pm 0.02$ & $0.41$ & $0.41$ & $0.65 \pm 0.06$ & $0.65$ & $0.66$ & $0.34 \pm 0.01$ & $0.34$ & $0.34$ \\
$3.0-4.0$ & $0.40 \pm 0.01$ & $0.40$ & $0.41$ & $0.48 \pm 0.04$ & $0.47$ & $0.49$ & $0.33 \pm 0.01$ & $0.33$ & $0.34$ \\
$4.0-5.0$ & $0.38 \pm 0.02$ & $0.38$ & $0.39$ & $0.52 \pm 0.05$ & $0.50$ & $0.54$ & $0.33 \pm 0.01$ & $0.32$ & $0.34$ \\
$5.0-6.0$ & $0.38 \pm 0.02$ & $0.37$ & $0.40$ & $0.45 \pm 0.06$ & $0.43$ & $0.48$ & $0.31 \pm 0.01$ & $0.29$ & $0.34$ \\
$6.0-8.0$ & $0.36 \pm 0.02$ & $0.34$ & $0.39$ & $0.31 \pm 0.04$ & $0.27$ & $0.35$ & $0.27 \pm 0.01$ & $0.24$ & $0.32$ \\
$~~8.0-10.0$ & $0.33 \pm 0.03$ & $0.29$ & $0.39$ & $0.34 \pm 0.05$ & $0.27$ & $0.45$ & $0.26 \pm 0.02$ & $0.20$ & $0.41$ \\
$10.0-12.0$ & $0.26 \pm 0.03$ & $0.20$ & $0.35$ & $0.23 \pm 0.05$ & $0.15$ & $0.40$ & $0.22 \pm 0.03$ & $0.13$ & $0.50$ \\
$12.0-15.0$ & $0.35 \pm 0.06$ & $0.23$ & $0.58$ & $0.18 \pm 0.05$ & $0.09$ & $0.47$ & $0.25 \pm 0.05$ & $-$ & $-$ \\
$15.0-18.0$ & $-$ & $-$ & $-$ & $0.23 \pm 0.14$ & $-$ & $-$ & $-$ & $-$ & $-$  \\
\bottomrule                                  
\end{tabular}
\tablefoot{In the table we list the abundance measurements in each radial bin and for each cluster, where: $Z_{\mathrm{joint}}$ are the values derived using the azimuthally averaged XRB parameters, shown in Fig. \ref{spec_prof} (right) as red dots, blue squares, and green stars; $Z_{\mathrm{low}}$ and $Z_{\mathrm{high}}$ are the profiles with the lowest and highest measured abundance values in the outermost radial bin (Fig. \ref{individual_prof}). }
\label{table:2}
\end{table*}

The temperature and abundance profiles of the three objects show typical characteristics of cool-core clusters. In particular, the temperature has a pronounced central drop due to radiative cooling in this region, while the metallicity profiles show a clear peak (reaching supersolar values in the case of A2589), likely associated with the brightest central galaxy (BCG). At intermediate radii, the three clusters reach peak temperatures of $\sim 3.5-4.0$ keV, typical for intermediate-mass objects, while outwards the observed temperature is below $2.5$ keV. Regarding the iron abundance, beyond $\sim 0.3~R_{500}$ the profiles of the three objects are flat at a level consistent with that of massive objects (\citealt{werner06,simionescu17,urban17,ghizzardi21}). In particular, in the outermost bins, we measured $0.35 \pm 0.06$, $0.23 \pm 0.14$, and $0.25 \pm 0.05$ Z$_{\odot}$ for MKW3s, A2589, and Hydra A, respectively. The question of whether the abundance profiles are flat in their outskirts at a similar level to massive ones is particularly relevant in the context of chemical enrichment processes in large structures, as we further discuss in Sect. \ref{flat_prof}. 

\subsection{Impact of individual XRB modelling}
\label{individ_meas}

We then characterised the soft X-ray background from the multiple circular regions placed around each cluster (Fig. \ref{clusters}) to obtain independent measurements of this background component. The aim of this test is to assess the effect of local XRB estimates, as may occur when azimuthal coverage of the outer regions of a cluster is not available, on the final cluster measurements. The background parameters measured from each region are shown in Figs. \ref{XRB_mkw3s}, \ref{XRB_a2589}, and \ref{XRB_hydra}, and their effect on the temperature and metallicity measurements is discussed in detail in Appendix \ref{app1}, with reference to Fig. \ref{individual_prof}. The total scatter (i.e. from minimum to maximum value) of the temperature and metallicity profiles due to the different XRB modelling is also shown in Fig. \ref{spec_prof} using shaded bands for the three objects. The upper and lower metallicity values obtained in each bin are also given in Table \ref{table:2} as $Z_{\mathrm{low}}$ and $Z_{\mathrm{high}}$, respectively.

As can be seen in Figs. \ref{XRB_mkw3s}, \ref{XRB_a2589}, and \ref{XRB_hydra}, the individual XRB measurements do not differ significantly from the azimuthally averaged values. While the individual differences are small, the combined effect of the XRB parameters has a substantial impact on the final cluster measurements. All profiles agree well up to $\sim 0.4~R_{500}$, as within this range the source emission dominates over the background. Indeed, in these regions, the shaded areas in Fig. \ref{spec_prof} are narrower than the statistical uncertainties in each bin. Moving towards the outskirts, the differences become more pronounced. In particular, temperatures in the outer regions can differ by up to $\sim50\%$ between the different estimates, though less so in A2589, likely due to the limited number of offset regions available, or the slightly narrower radial coverage considered in this study. Meanwhile, metallicity values at $\sim 0.8~R_{500}$ range from $0.1$ to $0.6$ Z$_{\odot}$. In these outer regions, the discrepancies exceed the statistical uncertainties of the azimuthally averaged profiles, although the individual measurements often remain consistent with the reference values within $3\sigma$. 

Finally, it is important to note that the observed discrepancies in the outskirts cannot be simply attributed to statistical fluctuations in the background spectra. As shown in Appendix \ref{app_flux}, the net fluxes measured from the background regions are, in most cases, statistically inconsistent with each other, indicating that distinct XRB modelling is required. Therefore, the different temperatures and abundances measured in the outer regions of MKW3s, A2589, and Hydra~A reflect a major source of systematic uncertainty arising from inaccurate XRB modelling, which can be effectively mitigated by adopting an azimuthal average around each cluster.


\section{Discussion}
\label{discussion}

\subsection{Investigating the variability}
\label{variability}

In the previous section, we showed that modelling the soft X-ray background from different regions of the sky leads to significant variations in the cluster parameters, especially in the outer regions of the three clusters. The reasons for this behaviour are captured in Fig. \ref{did_pic}. In poor galaxy clusters such as MKW3s, A2589, and Hydra A, temperatures and abundances are primarily derived from the Fe L-shell at $0.9-1.3$ keV in the ICM spectra. This is especially true in the outer regions, where the source emission is faint and the background contribution dominates the total signal. In such cases, the emission from the L-shell becomes comparable to that from the X-ray sky background components. Since these components also have emission peaks in a similar energy range, even small changes in their estimated parameters can significantly affect the reconstructed shape of the Fe L-shell and impact the results obtained\footnote{We note that potential systematics associated with the characterisation of the XMM-\textit{Newton} instrumental background may also affect measurements in the outer regions. While this is generally the case, we show in Appendix~\ref{app2} that the impact of instrumental background systematics on the selected clusters is secondary compared to that of the soft X-ray background, which remains the primary focus of this study.}. 
\begin{figure*}
            \centering
            \includegraphics[width = \textwidth]{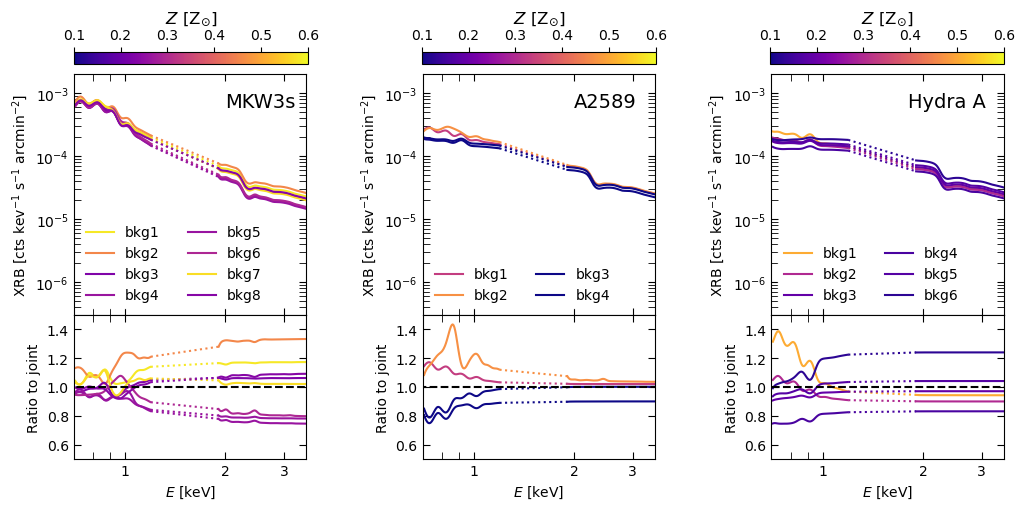}\\
            \caption{\footnotesize Top panels: Soft X-ray background models obtained from individual offset regions around each cluster (see Fig. \ref{clusters}). From left to right: MKW3s, A2589, and Hydra A. Each model results from the combination of the parameters shown in  Figs. \ref{XRB_mkw3s}, \ref{XRB_a2589}, and \ref{XRB_hydra}, and is colour-coded according to the metal abundance derived in the outermost bin for each cluster, using that particular parameterisation. Bottom panels: ratio of each individual XRB model to the azimuthally averaged model for each cluster. Models between 1.2 and 1.9 keV are indicated with dotted lines, as this energy range was excluded from the fitting procedure.}
            \label{fg_models}
\end{figure*}
\vspace{0.1cm}
\begin{figure*}
            \centering
            \includegraphics[width = \textwidth]{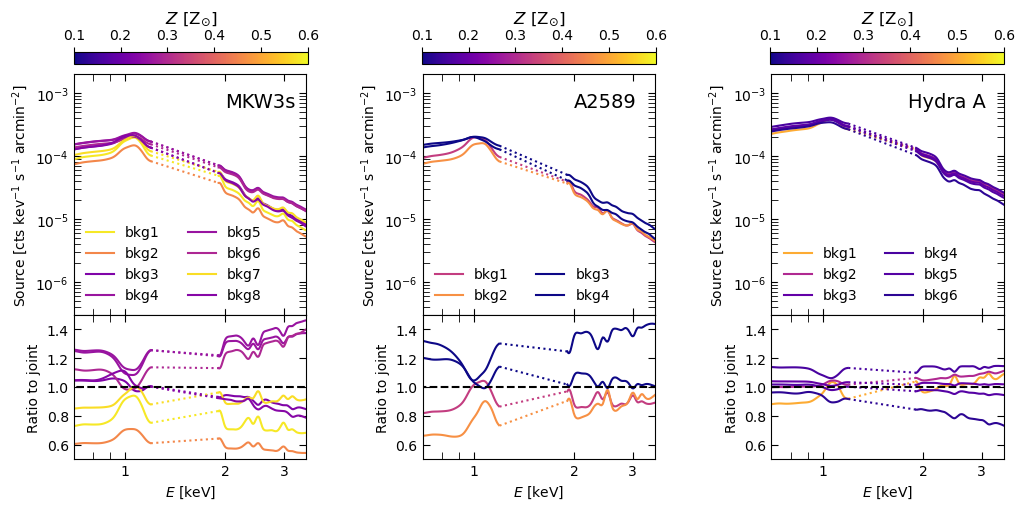}\\
            \caption{\footnotesize Same as Fig. \ref{fg_models}, but for the reconstructed source models.}
            \label{source_models}
\end{figure*}

Starting from the best-fitting parameters shown in Figs.~\ref{XRB_mkw3s}, \ref{XRB_a2589}, and \ref{XRB_hydra}, we derived the overall XRB models obtained from individual background regions and present them in the top panels of Fig.~\ref{fg_models}. To quantify the relative variations, these models are normalised to the corresponding azimuthally averaged XRB models in the bottom panels. The curves are colour-coded according to the iron abundance measured in the final radial bin of each cluster when adopting that particular XRB parameterisation, to highlight the impact of background modelling on the final abundance measurements. We find that estimating the XRB parameters from individual regions can introduce variations of up to $\sim20\%$ in the overall XRB models (see also discussion in Appendix \ref{app_flux}). These deviations appear to be correlated with the reconstructed iron abundance: specifically, an underestimated XRB model results in a lower measured metallicity, whereas a higher XRB level corresponds to a higher metallicity estimate. 

Naturally, these differences directly influence the shape and normalisation of the source emission models in the corresponding radial bins. This effect is illustrated in Fig.~\ref{source_models}, which shows the source models for the three clusters, adopting the same colour coding as in Fig.~\ref{fg_models}. As both the source and background models aim to reproduce the same observed spectra, underestimating the XRB level inevitably leads to an overestimate of the source model, and vice versa. This behaviour results in an anti-correlation between the source model normalisation and the derived iron abundance.
\begin{figure*}
    \centering
    \hspace{-0.2cm}\includegraphics[width = 0.45\textwidth]{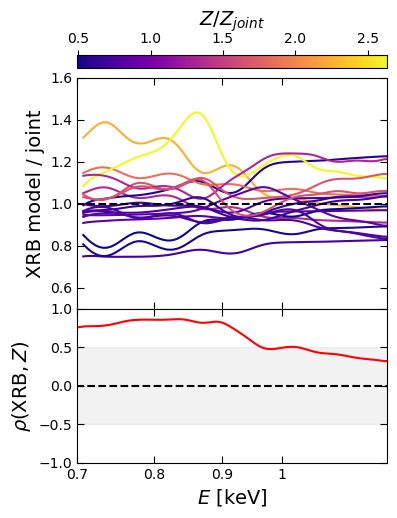} \hspace{0.5cm}
    \includegraphics[width = 0.45\textwidth]{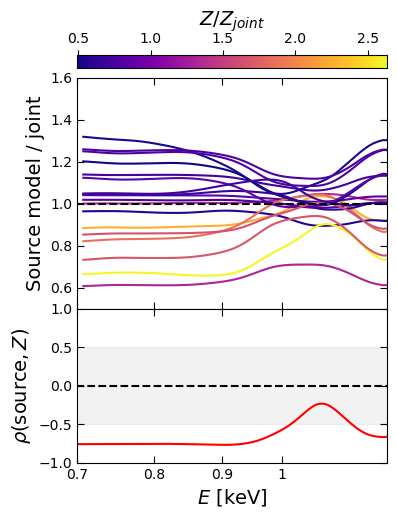}
    \caption{\footnotesize \footnotesize Left: the top panel shows the ratio of each XRB model to the corresponding azimuthally averaged one, for each cluster. The models are colour-coded according to the final $Z/Z_{\mathrm{joint}}$ values, where $Z$ is the iron abundance in the last bin of each cluster and $Z_{\mathrm{joint}}$ is the azimuthally averaged measurement. In the bottom panel, we plot the Pearson's $\rho$ parameter, which quantifies the correlation between XRB models and iron abundance (as plotted in the top panel), as a function of energy. Right: same as left, but for the reconstructed source models.}
    \label{all_models}
\end{figure*}

To further investigate the correlation between XRB and source models with the measured cluster metallicity, we combined the results from all three clusters and show them in Fig.~\ref{all_models} (top panels), normalised to their respective reference values. Differently from Figs.~\ref{fg_models} and \ref{source_models}, the colour coding here reflects the ratio $Z/Z_{\mathrm{joint}}$, where $Z$ is the metallicity derived using a specific XRB model, and $Z_{\mathrm{joint}}$ is the reference value obtained from the joint fit of the background spectra. This figure allows us to refine our considerations. As previously noted, background models that overestimate the reference lead to lower source model normalisations, which in turn yield higher metallicity values. Conversely, underestimated XRB levels result in higher source model normalisations, and thus lower metallicity estimates. However, it is important to emphasise that this relation is not linear: systematic effects are amplified such that $\sim 20\%$ differences in the XRB model can lead to more than $\sim 100\%$ variation in the reconstructed iron abundance.

We also computed Pearson's $\rho$ correlation coefficient to quantify the relationship between the XRB or source models and the reconstructed cluster metallicity, as a function of energy (Fig.~\ref{all_models}, bottom panels). The correlation is strongest between the XRB models and the measured metallicity in the soft $0.7-0.9$ keV band (Fig.~\ref{all_models}, left), gradually decreasing at higher energies where the cluster emission increasingly dominates over the background. This finding supports our choice to exclude the very soft $0.5-0.7$~keV band from the source fitting procedure: in that range, the XRB contribution becomes dominant, and even small variations in its modelling could introduce significant biases in the derived cluster properties. In Fig.~\ref{all_models} (right), we also show the anti-correlation between the source model normalisation and the reconstructed metallicity. This behaviour can be explained by the fact that the Fe L-shell region provides strong constraints with sufficient statistical power. As a result, the spectral fit tends to adjust the metallicity in this energy band to properly fit the data, regardless of the measured source normalisation. Consequently, a lower source model yields a higher metallicity, and vice versa. This interplay is most evident at $\sim 1$ keV, where the source reconstructions tend to converge with the reference model, and the correlation between source model and metallicity reaches a minimum.

\subsection{Flat iron abundance profiles}
\label{flat_prof}
As already mentioned, the outskirts of galaxy clusters offer an invaluable opportunity to study the processes that drive the chemical enrichment of the large-scale structure of the Universe. 

Beyond the core, hydrodynamic simulations that include non-gravitational processes have revealed a remarkable uniformity in iron abundance profiles across different systems \citep[e.g.][]{fabjan10,mccarty11}. This was originally interpreted as strong support for the `early enrichment' scenario, in which a combination of supernova explosions, stellar winds, and feedback from supermassive black holes played a key role in expelling metals from their host galaxies into the surrounding medium during, or even before, the formation of the hot ICM \citep[see, e.g.,][]{mernier18a,biffi18}. 

In the past decade, several studies have investigated the evolution with redshift of the metal abundance in cluster outskirts, finding little to no evolution from $z \sim 2$ to the present day (e.g. \citealt{ettori15, mcdonald16,mantz20,flores21}). \citet{molendi24} refined this framework by proposing that the metal enrichment in regions outside the core of galaxy clusters (the so-called `apex accretors') is dominated by \textit{ex-situ} processes. Specifically, they suggested that the observed iron mass in these regions was produced at $z \sim 2-3$ by halos with total masses of $\sim 10^{12}$ M$_{\odot}$, before being expelled into the intergalactic medium and later accreted onto the main cluster. This process is expected to operate similarly across all mass scales, naturally leading to a uniform iron abundance in the outer regions of all systems. Based on this model, the authors conclude that the enrichment efficiency has remained essentially constant from $z \sim 2$ to the present day. Observations of massive clusters support this general picture, with iron abundance profiles found to be relatively flat at similar values out to large radii \citep[e.g.][]{werner13, urban17, simionescu17, ghizzardi21}. However, as discussed previously, inconsistent results have been reported for galaxy groups, raising questions about whether this same enrichment history applies also for lower-mass systems. Our results for intermediate-mass objects allow us to shed further light on this issue.

Measuring iron abundances in the outskirts of intermediate-mass systems is feasible, although these are sensitive to systematic uncertainties. In particular, a major source of systematics in the outermost regions arises from inaccurate modelling of the XRB contamination. We have shown that local background regions (e.g. green circles in Fig. \ref{clusters}) cannot be reliably used to estimate the XRB parameters, as significant differences are observed at these scales (Appendix \ref{app_flux}). These variations propagate into the final cluster measurements, resulting in different abundance profiles in the outskirts, sometimes decreasing and some other times increasing with radius (see Sect. \ref{individ_meas} and Fig. \ref{individual_prof}). However, by applying an azimuthal characterisation of the XRB (Sect. \ref{azimuthal_aver}), we mitigated this source of uncertainty and obtained more reliable abundance profiles for the three clusters in our sample. The profiles remain nearly flat out to at least $\sim0.8~R_{500}$, consistent with those observed in more massive systems and supporting the idea that similar enrichment processes have occurred across the entire cluster population. Nevertheless, the large systematic uncertainties in the outermost regions still prevent us from excluding the possibility of declining iron-abundance profiles at large radii in these intermediate-mass systems.

The systematics identified in this work are expected to have an even greater impact on galaxy groups, where the X-ray emission is dominated by Fe L-shell lines. Therefore, more extensive studies that explicitly address these systematics in lower-mass systems are essential to determine whether the lower iron abundances sometimes observed in groups are real or simply the result of observational biases. If abundance levels consistent with those in more massive clusters are confirmed, this would provide strong support for the `early enrichment' scenario as the dominant mechanism of chemical enrichment across all mass scales. Conversely, persistent discrepancies would necessitate a re-evaluation of our current understanding of such processes.

\subsection{Combining X-ray and optical measurements}
\label{iron_yields}

\begin{table*}
\caption{\footnotesize Iron mass diffused in the ICM, stellar masses, and effective iron yields for MKW3s, A2589, and Hydra A.}
\centering 
\begin{tabular}{c|c|c|c|c|c|c}
    \toprule
    \toprule
    & \multicolumn{2}{c|}{$M_{\mathrm{Fe,500}}^{\mathrm{ICM}}$} & \multicolumn{2}{c|}{$M_{500}^{\mathrm{star}}$} & \multicolumn{2}{c}{$\mathcal{Y}_{\mathrm{Fe,500}}$} \\
    \cline{2-7}
    & Best fit & Systematics & Best fit & Systematics & Best fit & Systematics \\
    & ($10^{10}$ M$_{\odot}$) & ($10^{10}$ M$_{\odot}$) & ($10^{12}$ M$_{\odot}$) & ($10^{12}$ M$_{\odot}$) & (Z$_{\odot}$) & (Z$_{\odot}$) \\
    \midrule
    MKW3s & $1.28 \pm 0.06$ & $[+0.35,-0.20]$ & $2.86 \pm 0.44$ & $[+0.00,-1.14]$ & $2.68 \pm 0.34$ & $[+2.36, -0.33]$ \\
    A2589 & $0.92 \pm 0.09$ & $[+0.55,-0.14]$ & $2.20 \pm 0.63$ & $[+0.51,-0.51]$ & $2.54 \pm 0.64$ & $[+2.13, -0.62]$ \\
    Hydra A & $1.25 \pm 0.04$ & $[+0.46,-0.20]$ & $0.84 \pm 0.17$ & $[+0.22,-0.00]$ & $7.51 \pm 1.47$ & $[+0.13, -3.31]$ \\
    \bottomrule
\end{tabular}
\tablefoot{$M_{\mathrm{Fe,500}}^{\mathrm{ICM}}$ refers to the iron mass diffused in the ICM, computed within $R_{500}$. The systematic errors associated with these measurements originate from abundance profiles derived using individual XRB modelling, as described in Sect. \ref{iron mass}. The term $M_{500}^{\mathrm{star}}$ refers to stellar mass measurements within $R_{500}$, including both statistical and systematic errors, as detailed in Appendix \ref{sect_stellar}. Finally, we report the effective iron yields, combining the best-fit measurements and systematic errors of the previous two quantities, as discussed in Sect. \ref{sect_yields}.}
\label{table:3}
\end{table*}

In this section, we combine the X-ray measurements for the three clusters with those available at optical wavelengths, in order to obtain information on the total iron content within a given radius $R$, including the iron diffused in the ICM and that locked in stars and galaxies, as well as the total stellar mass that should have produced it. First, we provide measurements of the gas fractions of MKW3s, A2589, and Hydra A within $R_{500}$. We then calculate the iron mass diffused in the ICM of the three clusters, accounting for differences in metal abundance profiles resulting from individual XRB modelling (see Sect.~\ref{results}). Finally, we combine the iron masses with the available stellar masses to calculate their iron yields. The final measurements of the iron mass within the ICM, the stellar mass, and the iron yields of the three clusters in our sample are summarised in Table~\ref{table:3}. We note that uncertainties on the mass estimates were not propagated to the results, in order to ensure a consistent comparison with previous studies.

\subsubsection{Gas fraction calculation}
\label{gas fraction}
We calculated the total gas mass for each cluster in our sample through direct integration of the radial gas density profiles $\rho_{\mathrm{gas}}(r)$ within $R_{500}$, which are related to the electron number density $n_e(r)$ via $\rho_{\mathrm{gas}}(r) = \mu_e m_{\mathrm{H}} n_e(r)$. Here, $\mu_e = 1.11$ is the mean molecular weight per electron, and $m_\mathrm{H}$ is the atomic mass unit. The electron density profiles $n_e(r)$ were measured using the \textit{hydromass} code \citep{eckert22a}, through a non-parametric deprojection of the azimuthal median surface brightness profiles of the three clusters, extracted in the soft $0.7-1.2$ keV energy band. In this process, we included a radial-dependent conversion factor between count rate and emissivity, accounting for variations in temperature and metal abundance. Finally, by dividing the derived gas mass values by the corresponding total masses within $R_{500}$ (see Sect.~\ref{sect_mass}), we obtained constraints on the gas fractions, finding $f_{\mathrm{gas},500} = 0.095 \pm 0.001$, $0.075 \pm 0.002$, and $0.147 \pm 0.001$ for MKW3s, A2589, and Hydra A, respectively.

The gas fractions within $R_{500}$ measured for MKW3s and A2589 are typical for systems in the intermediate-mass range. In fact, they follow the parameterisation proposed by \citet{eckert21}, who compiled observational measurements spanning from galaxy groups to massive clusters (see their Fig. 7). The case of Hydra A is notably different, as it can be considered an outlier to this relation. Its gas fraction, $f_{\mathrm{gas},500}$, is close to the cosmological baryon fraction, an exceptional feature for a system of this mass.  As we will see in Sect.~\ref{sect_yields}, this has direct implications for the iron yield measured in Hydra A.

\subsubsection{Iron mass diffused in the ICM}
\label{iron mass}
Following \citet{degrandi04}, the ICM abundance profile $Z_{\mathrm{Fe}}(r)$ can be used to derive the total iron mass diffused in the ICM within $R_{500}$ as follows:
\begin{equation}
    M_{\mathrm{Fe,500}}^{\mathrm{ICM}} = 4 \pi A_{\mathrm{Fe}} m_{\mathrm{H}} Z_{n,\odot} \int_0^{R_{500}} Z_{\mathrm{Fe}}(r)\, n_{\mathrm{H}}(r)\, r^2\, \mathrm{d}r,
    \label{Eq_MFe}
\end{equation}
where $A_{\mathrm{Fe}}$ is the atomic weight of iron, $Z_{n,\odot} = 3.16 \times 10^{-5}$ is the solar abundance by number \citep{ghizzardi21}, and $n_{\mathrm{H}}(r)$ is the proton density profile of the cluster, obtained by assuming $n_e = 1.18\, n_{\mathrm{H}}$ \citep{asplund09}. Here, $Z_{\mathrm{Fe}}(r)$ is the three-dimensional iron abundance profile of the cluster, derived by deprojecting the abundance profile obtained from spectral fitting. For the deprojection, we employed the standard `onion-peeling' technique \citep{kriss1983, ettori02}, including a correction factor to account for emission from cluster regions beyond the outermost radial bin \citep[see][for details]{mclaughlin99, ghizzardi04}.

First, we deprojected the abundance profiles obtained using azimuthally averaged XRB parameters (i.e. our best-fit measurements; see Sect. \ref{results}) and used Eq. \ref{Eq_MFe} to compute the iron mass diffused in the ICM of MKW3s, A2589, and Hydra A within $R_{500}$. While these profiles provide the most accurate estimate of the iron masses in the three clusters, the results obtained using individual XRB measurements can be used to evaluate the potential effect of such systematic errors on this computation. For this reason, we also deprojected the upper and lower measured $Z(r)$ profiles for each cluster (Table \ref{table:2}), and computed the iron masses associated with these measurements. Since the deprojection is emission-weighted, we also used the associated spectral temperatures and normalisations for consistency. The differences between lower/upper iron mass measurements and the reference ones can be considered as the final systematic error resulting from the X-ray analysis. Our measurements of $M_{\mathrm{Fe,500}}^{\mathrm{ICM}}$ for MKW3s, A2589, and Hydra A are reported in Table \ref{table:3}, where we specify both statistical and systematic uncertainties. As defined, the systematic uncertainties likely represent an upper bound on the true systematics affecting the X-ray measurements. Nevertheless, as we will show, they still allow us to place meaningful constraints on the iron yields of the three clusters.

\subsubsection{Iron yields}
\label{sect_yields}
\begin{figure*}
            \centering
            \includegraphics[width = 0.6\textwidth]{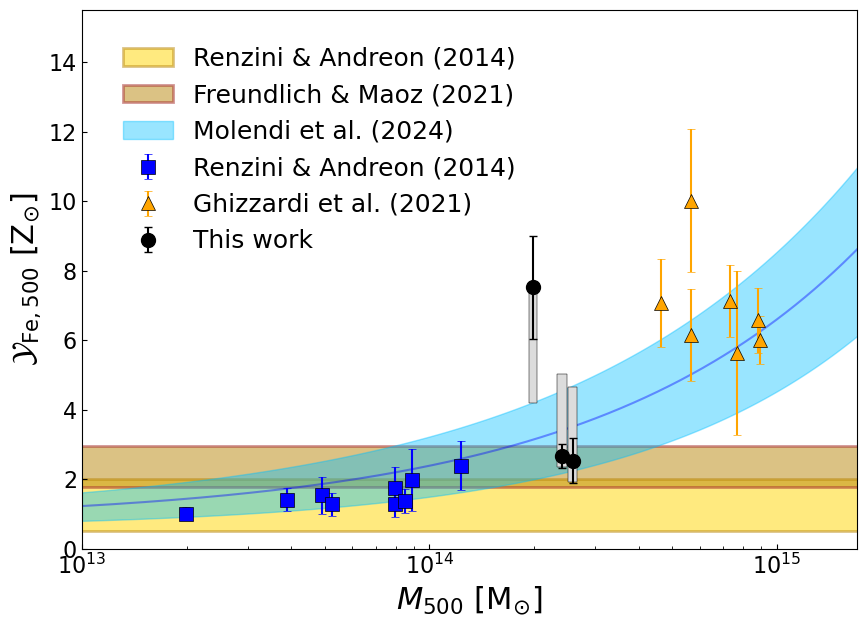}
            \caption{\footnotesize Iron yields. The measurements for MKW3s, A2589, and Hydra A, which include only statistical uncertainties, are shown in black. Vertical grey bands around these measurements indicate the systematic uncertainties, accounting for both iron and stellar mass estimates. Observational measurements from \citet{renzini14} and \citet{ghizzardi21-corrigendum} are shown as the blue boxes and orange triangles, respectively. Golden and brown regions show theoretical models based on supernovae \citep{renzini14,freundlich21}, while the light blue area is the parameterisation by \citet{molendi24}.}
            \label{YFe}
\end{figure*}

In this section, we further investigate the so-called `Fe Conundrum', i.e. the discrepancy observed between galaxy groups and massive clusters in terms of their effective iron yields \citep[][see Fig.~\ref{YFe}]{renzini14,ghizzardi21-corrigendum,gastaldello21}, which suggests the possibility of different efficiencies in iron production across these systems.

The enrichment model outlined in \citet{molendi24} provides new insights into this controversial topic as well. The authors attributed the reason for the `Fe Conundrum' to the different observed distributions of gas and stars in clusters and groups. Specifically, the star distribution as a function of radius is observed to be flatter in massive clusters than in galaxy groups (see their Eq. 15), while groups have a lower gas content than clusters, due to the stronger effect of AGN feedback at low mass scales \citep[see their Eq. 9;][]{eckert21}. Taking these observational effects into account, their parameterisation can reproduce the differences between the two extremes of the cluster mass scale (light blue area in Fig. \ref{YFe}). However, new measurements in the missing intermediate-mass range are necessary to either support this description or allow for alternative interpretations.

Following \citet{renzini14}, the effective iron yields within $R_{500}$ can be computed as: 
\begin{equation}
    \mathcal{Y}_{\mathrm{Fe,500}} = \frac{M_{\mathrm{Fe},500}^{\mathrm{ICM}} + M_{\mathrm{Fe},500}^{\mathrm{star}}}{M_{500}^{\mathrm{star}}(0)},
    \label{Eq_yield}
\end{equation}
where the numerator is the total iron mass within $R_{500}$ and the denominator is the stellar mass that should have produced it. More specifically, $M_{500}^{\mathrm{star}}(0)$ is the total mass that went into stars, reduced to the present observed value $M_{500}^{\mathrm{star}}$ by the return factor $r_0$, which quantifies stellar mass losses, i.e. $M_{500}^{\mathrm{star}}(0) = r_0 M_{500}^{\mathrm{star}}$. Following \citet{maraston05} and \citet{renzini14}, we adopted $r_0 = 1/0.58$. Regarding the iron content, $M_{\mathrm{Fe},500}^{\mathrm{ICM}}$ is the iron mass diffused in the ICM, computed as described in Sect. \ref{iron mass}, while  $M_{\mathrm{Fe},500}^{\mathrm{star}}$ indicates the iron mass still locked in stars. The latter is usually derived as the product of the stellar mass and an average stellar metallicity, e.g. solar. Following \citet{ghizzardi21}, we thus assumed that $M_{\mathrm{Fe},500}^{\mathrm{star}} = Z_{m,\odot}M_{500}^{\mathrm{star}}$, where $Z_{m,\odot} = 0.00124$ is the solar abundance by mass. Finally, we note that iron yields are typically reported in solar units.

As it is clear from Eq. \ref{Eq_yield}, a crucial point to address when performing iron-yield measurements is the selection of stellar masses, as these influence both $M_{\mathrm{Fe},500}^{\mathrm{star}}$ and $M_{500}^{\mathrm{star}}(0)$\footnote{The contribution of $M_{\mathrm{Fe},500}^{\mathrm{ICM}}$ to the total iron mass is a factor of $\sim5-10$ higher than the iron in stars $M_{\mathrm{Fe},500}^{\mathrm{star}}$ \citep[see e.g.][]{renzini14,ghizzardi21,molendi24}. Therefore, it is important to remember that the choice of the adopted stellar masses has a greater impact on the denominator term than on the numerator of Eq. \ref{Eq_yield}.}. We relied on publicly available measurements \citep[e.g. from][]{zhang11_corr,kravtsov18} for MKW3s and A2589. For Hydra A, we considered a measurement produced in the context of X-COP \citep{ghizzardi21}, following the procedure outlined in \citet{vanderburg15}. It is important to remember that, like metallicity measurements from X-rays, stellar masses can also be subject to large systematic errors. Therefore, we also derived a cross-calibration between the stellar mass values computed in the samples considered \citep[i.e.][]{zhang11_corr,kravtsov18,ghizzardi21}, based on clusters that are common to all samples, to obtain estimates of systematic errors associated with the measurements. The procedure for selecting and calibrating stellar masses, as used in our study, is detailed in Appendix \ref{sect_stellar}. The stellar mass values considered are also reported in Table \ref{table:3}, where we specify both statistical and systematic uncertainties.

We combined the stellar mass estimates with the measured iron mass in the ICM to calculate the iron yields of MKW3s, A2589, and Hydra A using Eq. \ref{Eq_yield}. For each cluster, we derived reference yield values using the best-fit measurements of both stellar and iron masses, including their respective statistical uncertainties. These are shown as black dots in Fig. \ref{YFe}. Given that systematic uncertainties are likely to dominate over statistical ones, we also propagated systematics associated with both the X-ray and optical measurements to define broader confidence intervals. Specifically, for each cluster, we computed an upper limit on the yield by combining the upper estimate of the iron mass with the lower estimate of the stellar mass, and a lower limit by reversing this combination. These systematic intervals are shown as grey shaded regions around the best-fit points in Fig. \ref{YFe}. It is important to note that these bands do not correspond to formal $68\%$ confidence intervals; rather, they indicate plausible ranges that reflect the most extreme variations in our iron yield measurements due to systematic uncertainties. The final results for MKW3s, A2589 and Hydra A, which account for both statistical and systematic uncertainties, are reported in Table  \ref{table:3}.

The values measured for the three clusters lie between those for galaxy groups and massive clusters. Notably, the yields for MKW3s and A2589 are around $\sim 2.5$ Z$_{\odot}$, consistent with the available measurements at the scale of galaxy groups \citep{renzini14}. Even when accounting for systematic uncertainties in iron and stellar mass measurements, these yields are in excellent agreement with the parameterisation proposed by \citet{molendi24}, but only marginally consistent with predictions based on supernova explosions \citep{renzini14,freundlich21}. In contrast, Hydra A displays a significantly higher yield of $\sim 7.5$ Z$_{\odot}$, comparable to those measured in massive clusters \citep{ghizzardi21-corrigendum}. This result is not entirely unexpected, as the unusually high gas content of Hydra A makes it an outlier in the $f_{\mathrm{gas},500}$–$M_{500}$ relation (see Sect.~\ref{gas fraction}). Nevertheless, once systematic uncertainties are considered, particularly those associated with stellar mass estimates, the iron yield of Hydra A may decrease significantly, thus becoming consistent with values found in lower-mass systems.

This study marks the first step towards a comprehensive investigation of the intermediate-mass regime in terms of iron yields, a regime that remains still largely unexplored. Future work based on larger, more homogeneous samples will be crucial in confirming these findings and providing a more comprehensive understanding of the iron enrichment process across the full mass scale of galaxy clusters.


\section{Summary and conclusions}
\label{conclusion}

This study builds on recent findings in the literature that suggest potential differences between galaxy groups and massive clusters, particularly in terms of the shape of their iron abundance profiles at large radii and the corresponding iron yields. In this context, clusters with masses in the range $M_{500} \simeq 1.5-3.5 \times 10^{14}$ M$_{\odot}$ are particularly well suited to investigating these issues, as they occupy the transition region between the two populations. However, the metal content of these intermediate-mass systems, especially in the outer regions, remains largely unexplored. Our analysis focuses on a pilot sample of three poor clusters, i.e. MKW3s, A2589, and Hydra A, with total masses of $M_{500} \simeq 2.0-2.5 \times 10^{14}$ M$_{\odot}$, with the aim of providing a detailed characterisation of their chemical properties at large radii and beginning to populate the intermediate-mass range of the cluster population.

Taking advantage of the extensive azimuthal coverage from XMM-\textit{Newton} observations available for the three clusters (Fig. \ref{clusters}), we assessed the impact of soft X-ray background (XRB) modelling on the derived iron abundance profiles. Our main findings can be summarised as follows:
\begin{itemize}
    \item modelling the XRB using only local regions of the sky near the cluster, rather than the full azimuthal coverage, introduces systematic uncertainties in the abundance measurements that exceed the statistical errors (Fig. \ref{individual_prof}). In particular, when XRB parameters are fixed at their best-fitting value and applied directly to the source fits, even modest variations ($\sim 20\%$) in the reconstructed XRB level can result in substantial differences ($\sim 100\%$) in the measured iron abundances at large radii of intermediate-mass systems (Fig. \ref{all_models});
    \item an azimuthal characterisation of the XRB yields more robust metallicity profiles for all three clusters. These profiles remain flat beyond the core regions, with iron abundances converging to $\sim 0.3$ Z$_{\odot}$ (Fig.~\ref{spec_prof}), in good agreement with the values observed in massive systems \citep[e.g.][]{werner13,urban17,ghizzardi21}. This result further supports the hypothesis of a common chemical enrichment history across the entire population of galaxy clusters, although the large systematic uncertainties in the cluster outskirts prevent us from drawing definitive conclusions.
\end{itemize}

We used the measured iron abundance profiles of MKW3s, A2589, and Hydra A to estimate the total iron mass diffused in their ICM. By combining these values with available stellar mass estimates, we derived the effective iron yields within $R_{500}$ for the three clusters (Fig. \ref{YFe}). Our main findings are summarised below:
\begin{itemize}
    \item the best-fit iron yields are $\mathcal{Y}_{\mathrm{Fe,500}} = 2.68 \pm 0.34$, $2.54 \pm 0.64$, and $7.51 \pm 1.47$ Z$_{\odot}$ for MKW3s, A2589, and Hydra A, respectively. The values for MKW3s and A2589 are consistent with those typically measured in galaxy groups \citep{renzini14}, while the yield for Hydra A is in line with measurements in massive clusters \citep{ghizzardi21-corrigendum};
    \item when accounting for systematic uncertainties, particularly those related to stellar mass values, the yields of the three clusters may converge towards similar values, intermediate between groups and massive clusters;  
    \item the measured yields are only marginally consistent with theoretical models in the literature based on supernova explosions \citep{renzini14,freundlich21}. Conversely, they are more compatible with the model proposed by \citet{molendi24}, discussing the enrichment history of clusters and groups.
\end{itemize}

In conclusion, reliable abundance measurements in the outer regions of intermediate-mass objects are possible, provided that systematic uncertainties are properly addressed. We have demonstrated that full azimuthal coverage is necessary, not only to ensure complete sampling of the ICM, but also to achieve an accurate characterisation of the local X-ray sky background. In particular, with a careful modelling of the XRB contamination, it is possible to obtain robust abundance measurements out to at least $\sim 0.85~R_{500}$ even in clusters within this mass range. The systematics identified in this work are expected to have an even greater impact in galaxy groups, where the X-ray emission is dominated by Fe L-shell lines. As such, dedicated studies focusing on lower mass systems are essential to accurately constrain the metal content within these systems.

As Fig.~\ref{YFe} shows, the intermediate-mass regime remains largely unexplored in terms of iron yields. We aim to extend this analysis to additional systems in this mass range, building a sample that is comparable in size to those available at lower and higher masses. While this work outlines a strategy to quantify X-ray-related systematics, equal attention must also be given to uncertainties in stellar mass estimates. For instance, the existing results on iron yields are based on heterogeneous samples that have been selected and analysed in different ways. In this sense, well-defined and minimally biased samples with high-quality X-ray observations and homogeneous, accurate stellar mass measurements are needed to derive final constraints on this topic.

Finally, it is worth noting that we are entering the era of space micro-calorimeters, currently on board XRISM \citep{tashiro18,guainazzi18}, and planned as a key component of the future \textit{NewAthena} mission \citep{cruise25,peille25}. These instruments will significantly reduce systematics related to limited spectral resolution, particularly in the Fe L-shell and K$\alpha$ regions, thereby improving the robustness of iron abundance measurements in both galaxy clusters and groups. At the same time, new instruments operating at other wavelengths will also advance the research in this area. For example, the high sensitivity of \textit{Euclid} \citep{laureijs11} will enable improved stellar mass measurements by detecting even the faint light of diffused stars in clusters \citep[e.g.][]{atek24,kluge24,ellien25}.


\begin{acknowledgement}
    We thank the anonymous referee for useful comments, which have improved our paper. This work is based on observations obtained with XMM-\textit{Newton}, an ESA science mission with instruments and contributions directly funded by ESA Member States and NASA. G.R. acknowledges D. Eckert and C. Grillo for useful discussions that improved the quality of the work. We acknowledge financial contribution from the INAF GO Programme "Iron in intermediate mass galaxy clusters" (1.05.23.05.16).  I.B. acknowledges the financial contribution from the contracts Prin-MUR 2022, supported by Next Generation EU (n.20227RNLY3; The concordance cosmological model: stress-tests with galaxy clusters). LL acknowledges support from INAF grant 1.05.12.04.01.
\end{acknowledgement}

\bibliographystyle{aa} 
\bibliography{biblio} 
\begin{appendix}

\section{Individual XRB parameters and impact on cluster measurements}
\label{app1}

In this section, we present the best-fitting measurements of the soft X-ray background parameters. These were obtained either through a joint fit of the spectra extracted from all the offset regions shown in Fig. \ref{clusters} or by considering each region individually. We also show the effect that each parameterisation has on the temperature and metallicity profiles of the three clusters.

\begin{figure*}
            \centering
            \includegraphics[width = 0.955\textwidth]{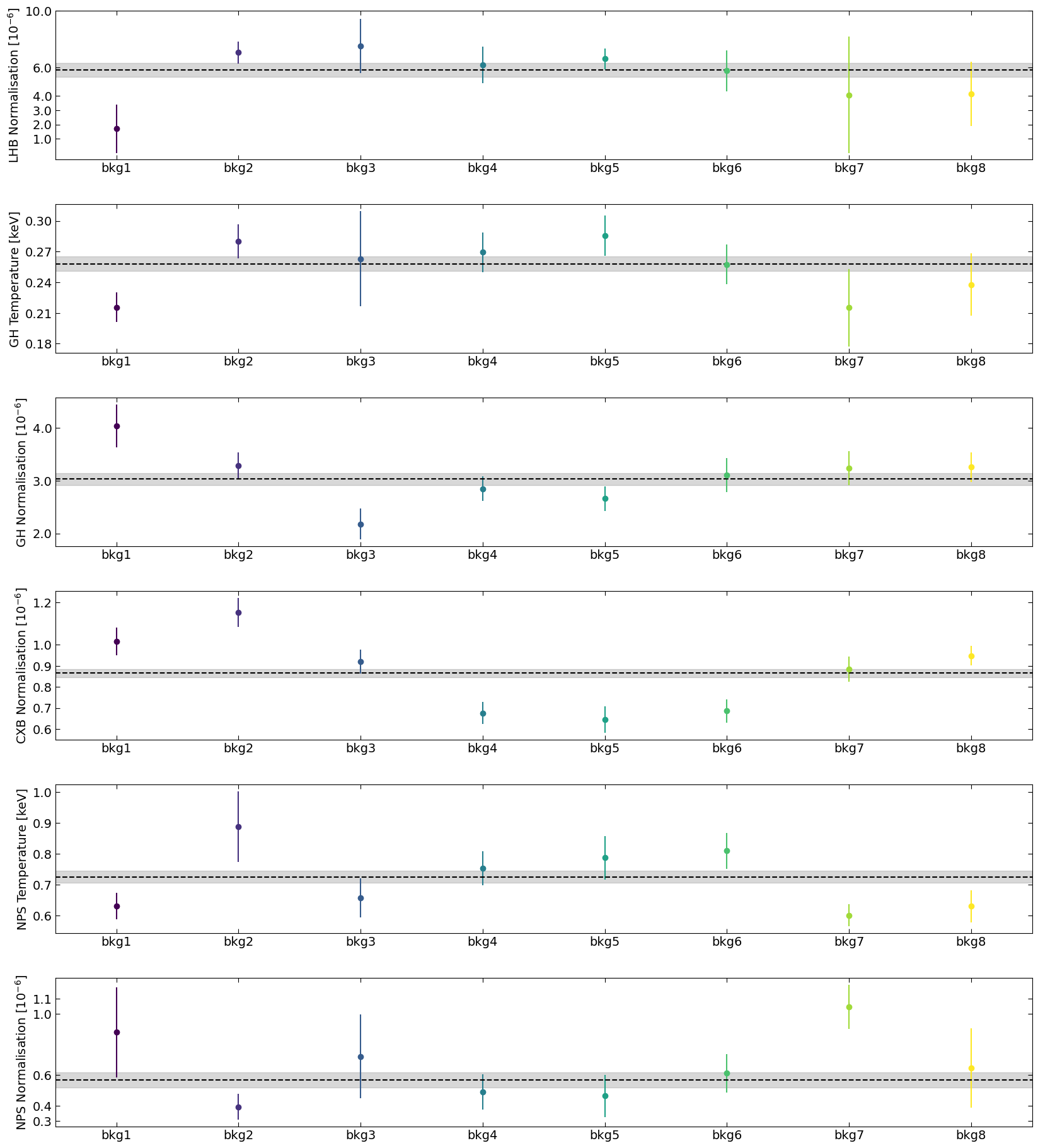}
            \caption{\footnotesize Best-fitting parameters of the soft X-ray background measured from the offset regions around MKW3s (Fig. \ref{clusters}, top). The azimuthally averaged measurements are shown as black dashed lines, and their statistical uncertainties are indicated by grey bands.}
            \label{XRB_mkw3s}
\end{figure*}

\begin{figure*}
            \centering
            \includegraphics[width = 0.94\textwidth]{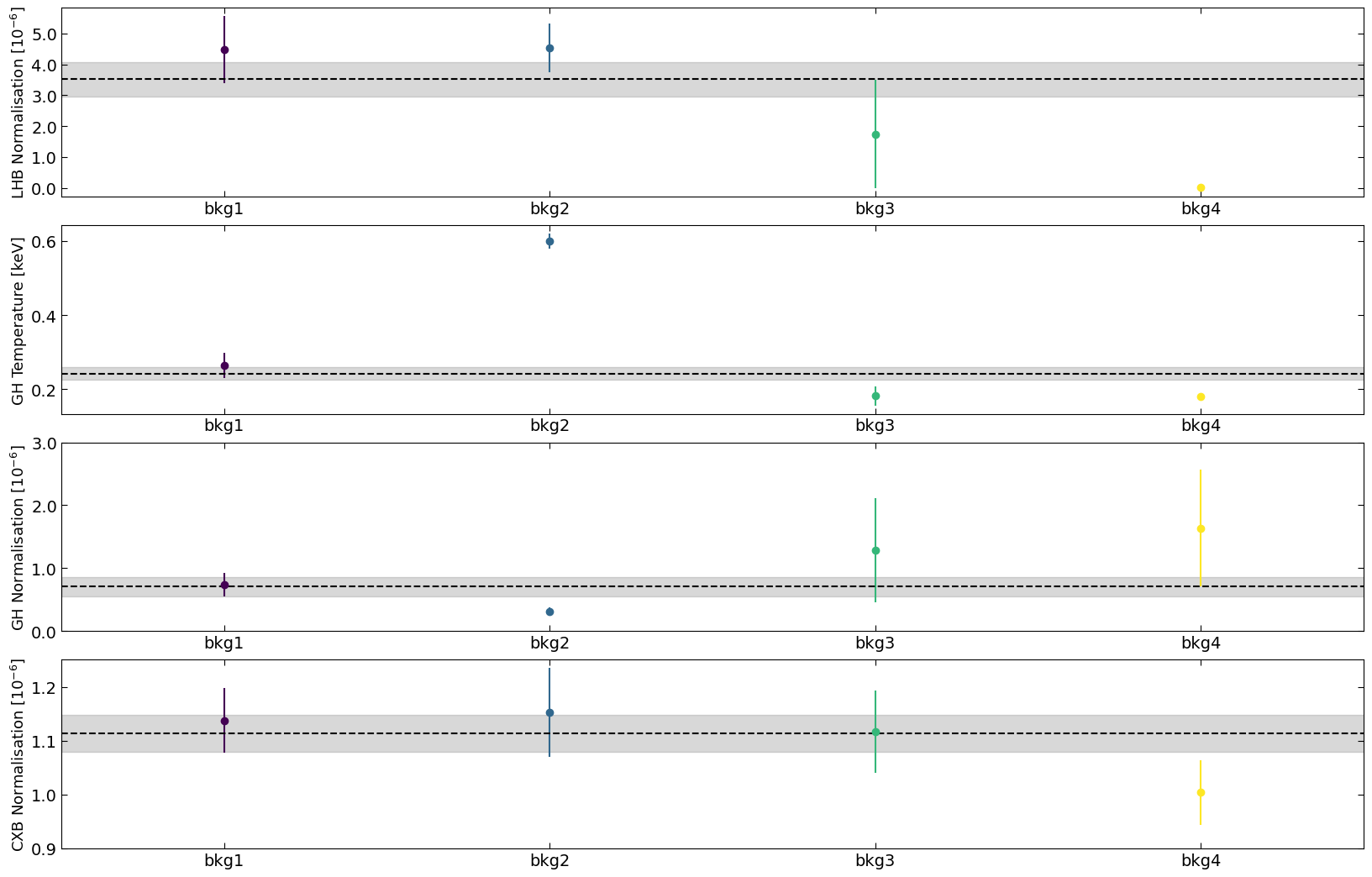}
            \caption{\footnotesize Same as Fig. \ref{XRB_mkw3s}, but for A2589 (Fig. \ref{clusters}, centre).}
            \label{XRB_a2589}
\end{figure*}

\begin{figure*}
            \centering
            \includegraphics[width = 0.96\textwidth]{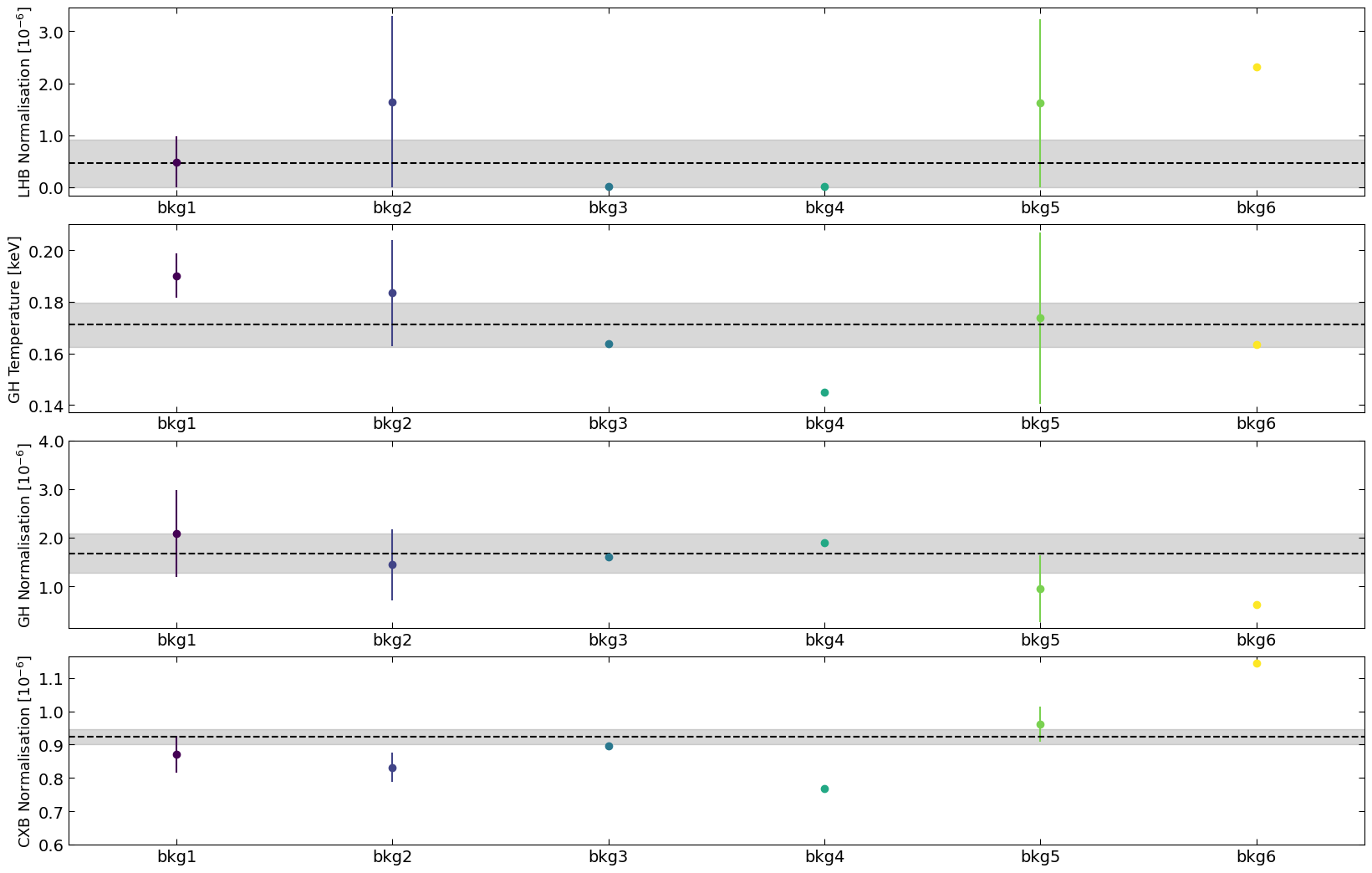}
            \caption{\footnotesize Same as Fig. \ref{XRB_mkw3s}, but for Hydra A (Fig. \ref{clusters}, bottom).}
            \label{XRB_hydra}
\end{figure*}

Figures \ref{XRB_mkw3s}, \ref{XRB_a2589}, and \ref{XRB_hydra} show the best-fitting measurements of the XRB parameters for MKW3s, A2589 and Hydra A, respectively. The measurements obtained from a joint fit of the offset regions are shown as dashed black lines, and the grey shaded regions indicate their statistical errors. In all cases, these parameters are well constrained. Superimposed on these measurements are the parameter measurements obtained from each region independently. In most cases, these measurements are not statistically discrepant from the reference azimuthal values, nor are any particular azimuth trends apparent. The only potential exception is the CXB measured for MKW3s; however, it is unclear whether this trend is real or arises from the combination of the other parameters.  Notably, the variability for MKW3s may be higher than for other clusters as it is projected onto the north polar spur. However, it is interesting to note that, although the parameters themselves are not particularly different, their combinations often are, as shown by the overall models discussed in Sect. \ref{variability}. Finally, we note that, in some cases (i.e. for Hydra A and A2589 only, where the background level is lower), the statistics did not permit the calculation of the error on certain XRB parameters. Nevertheless, this has no impact on our final results, as the background parameters are fixed at their best-fitting values during the spectral fitting of the source emission.

\begin{figure*}[h!]
            \centering
            \includegraphics[width = 0.88\textwidth]{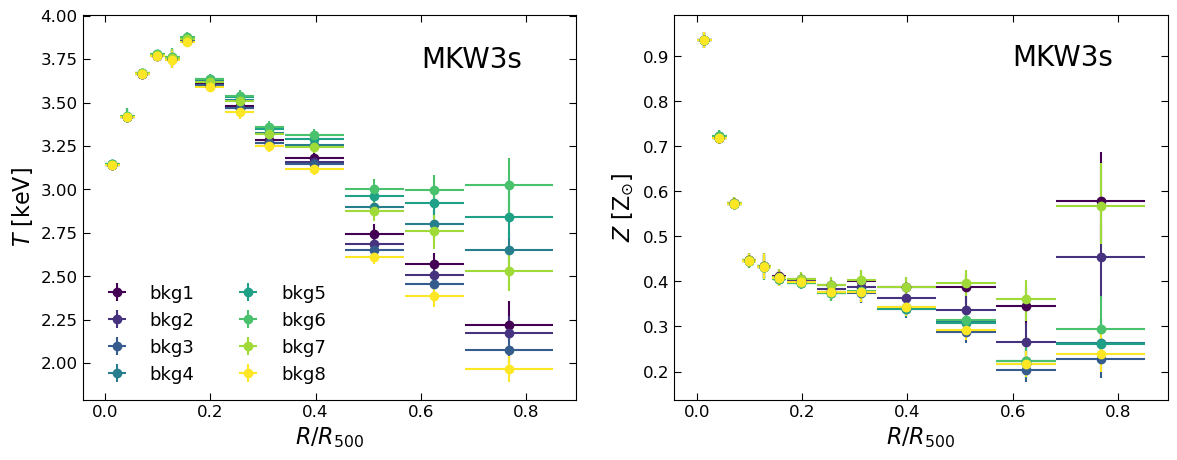}\\
            \includegraphics[width = 0.89\textwidth]{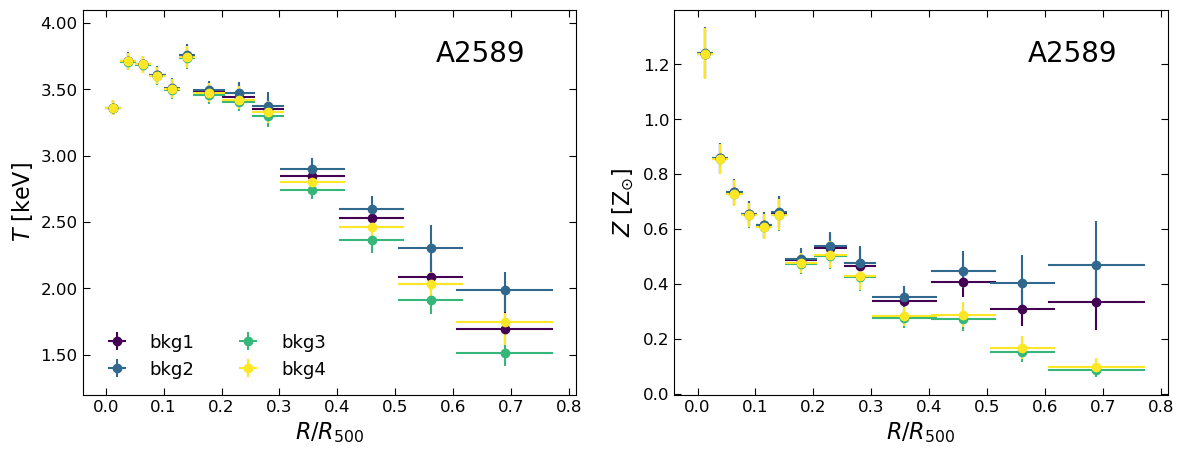}\\
            \includegraphics[width = 0.89\textwidth]{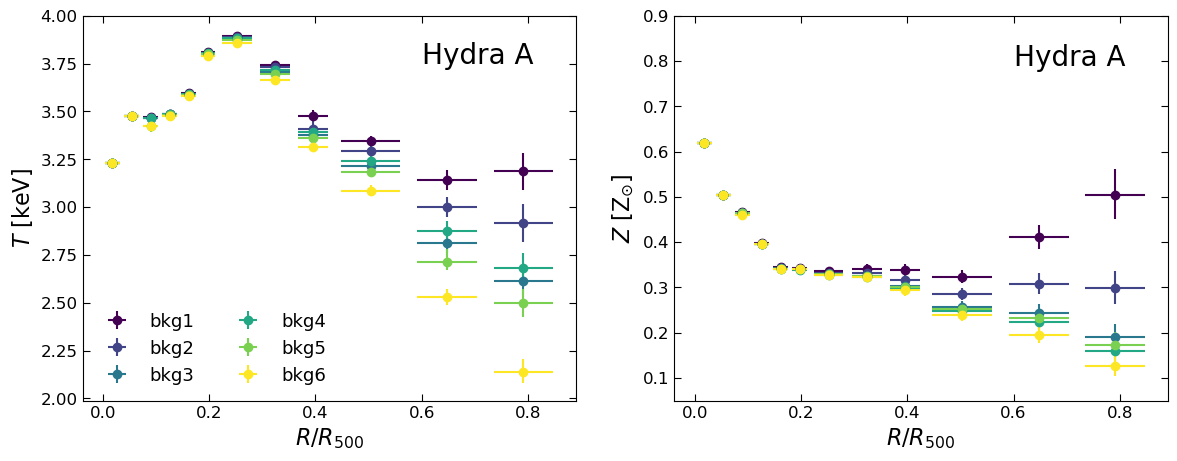}
            \caption{\footnotesize Temperature (left) and iron abundance (right) profiles obtained using different XRB models. The results for MKW3s, A2589 and Hydra A are shown in the top, middle and bottom panels, respectively.}
            \label{individual_prof}
\end{figure*}

As discussed in Sects. \ref{results} and \ref{variability}, using different values of the XRB parameters leads to variations in the reconstructed temperature and metallicity profiles of the three clusters. Figure \ref{individual_prof} provides a more detailed view of the impact of various XRB models on cluster temperatures (left panels) and abundances (right panels). The results for MKW3s, A2589 and Hydra A are shown in the top, middle and bottom panels, respectively. As can be seen from the figure, the variations in the temperature and metallicity profiles are small up to intermediate radii (i.e. $\sim 0.4~R_{500}$). Beyond this radius, however, the differences between the profiles increase significantly and exceed the statistical errors. These are systematic variations at each radius, which simply increase in the outer regions, where the background contribution to the source is larger. The total dispersions, given by the difference between the maximum and minimum profiles in each panel, are also plotted in Fig. \ref{spec_prof} as shaded areas. We mention that XRB modelling derived from spectra extracted from the same offset can, in some cases, lead to similar temperature and metallicity results, although there are some exceptions. 

\section{Flux measurements from individual XRB regions}
\label{app_flux}

We have shown that modelling the XRB spectral properties from local regions (e.g. green circles in Fig. \ref{clusters}) can lead to significant variations in the outskirts of intermediate-mass systems, whereas an azimuthal characterisation helps mitigate this source of systematic uncertainty. However, since the XRB properties were fixed at their best-fitting values when deriving the final cluster measurements, one may question whether the observed discrepancies simply reflect statistical fluctuations of the same XRB spectrum, rather than indicating a genuine source of systematics.

\begin{figure*}
            \centering
            \includegraphics[width = 0.92\textwidth]{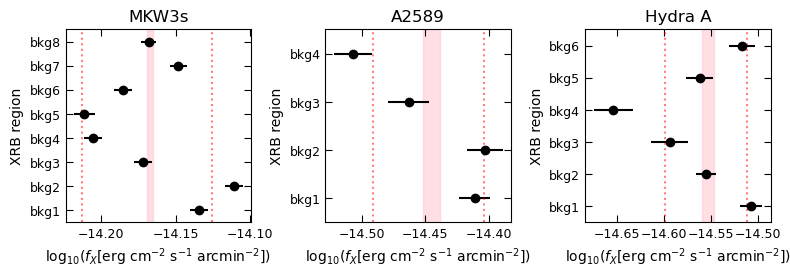}
            \caption{\footnotesize X-ray fluxes in the $0.5-2.0$ keV energy band measured from individual background regions (green circles in Fig. \ref{clusters}). The shaded bands indicate the fluxes obtained from the joint fit of all background regions, with the dotted vertical lines marking a $10\%$ variation from these values.}
            \label{flux_xrb}
\end{figure*}

To address this, we included the \textit{cflux}\footnote{\href{https://heasarc.gsfc.nasa.gov/docs/software/xspec/manual/node303.html}{https://heasarc.gsfc.nasa.gov/docs/software/xspec/manual/node303.html}.} component in our XRB models and measured the net XRB flux per square arcminute in the soft $0.5-2.0$ keV band, where the XRB contamination is strongest. These measurements are shown in Fig. \ref{flux_xrb} for each background region, along with the results from the joint fit of all regions (pink shaded area). As already noted, MKW3s shows higher XRB fluxes than the other two systems, as it lies on the NPS. However, all three clusters exhibit azimuthal variations of up to $\sim10-15\%$ relative to the values obtained from the joint fit (see also discussion in Sect. \ref{variability}), in line with previous results in the literature exploring the substantial variability of this background component \citep[e.g.][]{lumb02,bonamente05,henley13}. Notably, the individual flux measurements in Fig. \ref{flux_xrb} are, in most cases, statistically inconsistent with each other, suggesting that, on scales of $\sim50-80~\mathrm{arcmin}^2$, we are sampling independent realisations of the XRB emission that therefore require distinct modelling.

This analysis confirms that, when the XRB parameters are fixed and applied directly to the source regions, individual background regions cannot be reliably used to estimate them, as significant variations are observed on these spatial scales. Therefore, with this approach, full azimuthal coverage of the cluster outskirts is required to avoid unwanted systematic uncertainties in the outer regions of intermediate-mass systems.

\section{Impact of XMM-\textit{Newton} instrumental background}
\label{app2}

The focus of the present paper was primarily on the impact of the soft X-ray background on cluster measurements. This is a crucial aspect, since the emission peaks of the XRB components lie in the softer part of the ICM spectrum ($E \lesssim 1$ keV; see Fig. \ref{did_pic}), close to the cluster's Fe L-shell emission. Consequently, variations in the normalisation and spectral shape of these components significantly affect temperature and metallicity measurements for objects with temperatures $T \lesssim 2-3$ keV (Sect. \ref{variability}). However, the XMM-\textit{Newton} instrumental background component also impacts the measurements, particularly in the outer regions, where this is the dominant contribution to the total emission. Based on qualitative considerations, systematics related to this background component are expected to be more significant for massive objects, for which temperature and metallicity measurements are based on the exponential cut-off of the \textit{bremsstrahlung} emission, and on the K$\alpha$ line at $\sim 6.7$ keV. In this section, however, we quantify the potential impact of systematics related to this background component on the temperature and metallicity measurements for MKW3s, A2589, and Hydra A.

The instrumental background for each detector is modelled using a power law and a set of Gaussian models, that account for the fluorescence lines. Over the past few decades, extensive work has been dedicated to characterising this background component, and today, the systematics have been reduced to a few percent \citep{leccardi08a,salvetti17,marelli17,marelli21,gastaldello22,rossetti24}. In particular, in their description of the spectral analysis pipeline for CHEX-MATE galaxy clusters, \citet{rossetti24} report typical uncertainties of $\sim 2\%$ on the normalisation of the background model for MOS detectors, and of $\sim 6\%$ for the pn models. 

We assess the impact of these uncertainties on the measurements of the three clusters considered in this study. Specifically, we focus on the final radial bin of each cluster, where the impact of the background is most significant. We evaluate two extreme scenarios:
\begin{itemize}
    \item first, all background model normalisations related to MOS detectors were decreased by $2\%$, while those related to pn were decreased by $6\%$;
    \item second, we increased all MOS normalisations by $2\%$ and all pn normalisations by $6\%$.
\end{itemize}
The first case simulates the effect of a potential underestimation of the particle background, while the second case represents a potential overestimation. However, it is important to consider the above scenarios are upper limits of the true impact of background-related systematics. This assumes that every background component is incorrectly estimated in the same direction and by the maximum plausible uncertainty, which is an extremely unlikely condition. Nevertheless, this exercise allows us to derive meaningful conclusions about the influence of the instrumental background on our final measurements.

\begin{figure*}
            \centering
            \includegraphics[width = 0.9\textwidth]{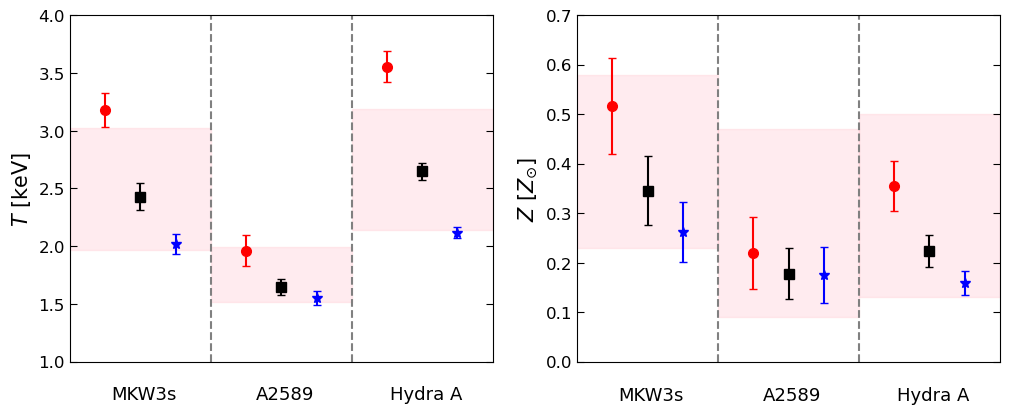}
            \caption{\footnotesize Impact of the XMM-\textit{Newton} instrumental background on the derived temperature (left) and iron abundance (right) in the outer regions of the clusters. The black squares show the reference measurements with no systematics considered. The red dots were derived using a lower instrumental background, while the blue stars were derived using an overestimate of this component, as described in the text. The pink shaded areas represent the dispersions associated with different XRB parameters within the same radial bin.}
            \label{inst}
\end{figure*}

The results of this test are shown in Fig. \ref{inst}, which illustrates the impact on cluster temperature (left) and iron abundance (right). The red points are the measurements obtained when assuming a lower instrumental background, and the blue points show the impact of an overestimated background. The black points represent the nominal measurements, which do not include systematics. As can be seen in the figure, changing the normalisation of the instrumental background clearly affects the measured temperature: decreasing the background contribution alters the spectral shape of the source model, resulting in a higher reconstructed temperature. Conversely, increasing the background results in a lower measured temperature. This directly affects the metallicity: due to the correlation between temperature and metallicity, configurations with a lower instrumental background tend to overestimate the metallicity and vice versa. Although temperature variations are significant, metallicity values often remain consistent with nominal measurements within the error bars.

Finally, Figure \ref{inst} also illustrates a comparison of variations due to under/overestimates of the XMM-\textit{Newton} instrumental background with those from different soft X-ray background models, within the same cluster bin (see Appendix \ref{app1}). The impact of the latter is indicated by the shaded pink areas. We find that the differences induced by the instrumental background are smaller than those associated with the different XRB models for iron abundances. As mentioned above, it is also important to note that these estimates are based on highly unlikely conditions and therefore represent an upper limit on the impact of the instrumental background on final cluster measurements. Therefore, we conclude that, within the regime explored in this study, the impact of systematics related to the instrumental background is smaller than that of the XRB, which remains the focus of the present work.

\section{Stellar masses}
\label{sect_stellar}
We searched for available stellar mass measurements of the three clusters in our sample that could be used to compute their effective iron yields (Sect. \ref{sect_yields}). For A2589, we identified a single value reported in the literature by \citet{zhang11_corr}. For MKW3s, two measurements were available, taken from \citet{zhang11_corr} and \citet{kravtsov18}; in this case, we adopted the weighted average of the two as the reference value for the iron yield computation. Finally, for Hydra A we used a stellar mass measurement produced in the framework of the X-COP project. We note that these observational measurements may rely on different assumptions for the initial mass function (IMF), and may also have been computed using different values of $R_{500}$. As detailed below for each cluster, we therefore applied basic corrections to the above measurements and derived stellar mass values under consistent conditions. In particular, we adopted \citet{chabrier03} as the reference IMF.

As discussed in Sect.~\ref{sect_yields}, we also derived a cross-calibration between the stellar mass measurements reported in the above papers, based on clusters that are common to all samples. Specifically, using five clusters (A85, A1795, A2029, A2142, and Zw1215), we found that the stellar masses reported by \citet{zhang11_corr} are, on average, $\sim30\%$ higher than the corresponding X-COP values presented in \citet{ghizzardi21} and computed following the procedure outlined in \citet{vanderburg15}. Conversely, considering the same clusters (excluding Zw1215, which is not included in either of those samples), we found that the estimates by \citet{kravtsov18} are, on average, $\sim60\%$ higher than the X-COP values. We used these systematic offsets to define confidence intervals for the stellar mass estimates of each cluster, and included them into the computation of the iron yields, as described in Sect.~\ref{sect_yields}. 

We note that since the measurements for MKW3s and Hydra A were computed by \citet{kravtsov18} and using the method of \citet{vanderburg15}, respectively, they already lie at opposite ends of this cross-calibration scale. As a result, the systematics defined for these two clusters are asymmetric, i.e. towards lower values for MKW3s and towards higher values for Hydra A, as detailed below. This asymmetry directly affects the shape of the systematic bars associated with their respective iron yields, as shown in Fig.~\ref{YFe}.

\subsection*{MKW3s}
Two values of stellar mass measurements are reported in the literature for MKW3s \citep{zhang11_corr,kravtsov18}, which need to be rescaled, since they use different IMF and $M_{500}$ (and so $R_{500}$) assumptions. Hence:
\begin{itemize}
    \item \citet{zhang11_corr} report $M_{500}^{\mathrm{star}} = (3.90 \pm 0.43)  \times 10^{12}$ M$_{\odot}$ as the stellar mass of the cluster, using a \citet{salpeter55} IMF and $M_{500} \simeq 1.45 \times 10^{14}$ M$_{\odot}$ for the total mass. First, we divided their measured mass value by $2$ \citep{chiu18}, accounting for the conversion from the \citet{salpeter55} to the \citet{chabrier03} IMF. Then, we multiplied it by $1.66$, to match their $M_{500}$ value to our measurement, thus assuming constant stellar fractions within the two $R_{500}$ values. In this conversion, we also included the uncertainties related to the $M_{500}$ measurements. The final stellar mass considered is: $M_{500}^{\mathrm{star}} = (3.24 \pm 0.87) \times 10^{12}$ M$_{\odot}$. 
    \item \citet{kravtsov18} report separate contributions to the stellar mass, both associated with the BCG and not included in it, for a total value of $M_{500}^{\mathrm{star}} = (2.68 \pm 0.49) \times 10^{12}$ M$_{\odot}$. They adopted a \citet{chabrier03} IMF and $M_{500} \simeq 2.35 \times 10^{14}$ M$_{\odot}$ for the total mass, which is slightly lower than our measured value. We thus simply multiplied their stellar mass by $1.03$ to account for the different $M_{500}$ (thus, $R_{500}$) measurement. As before, statistical uncertainties related to total mass measurements were included in the conversion to calculate the final error. The final considered stellar mass is then: $M_{500}^{\mathrm{star}} = (2.76 \pm 0.50) \times 10^{12}$ M$_{\odot}$.
\end{itemize}

Starting from the two available measurements, we adopted the weighted average as the reference stellar mass for MKW3s. In addition, based on the cross-calibration between \citet{zhang11_corr}, \citet{kravtsov18}, and \citet{ghizzardi21} discussed above, we defined a potential lower limit on the stellar mass by scaling the value reported by \citet{kravtsov18} down by a factor of 1.6. No upper limit could be established, as both \citet{zhang11_corr} and \citet{kravtsov18} are already included in the reference measurement. Therefore, accounting for both statistical and systematic uncertainties, the final stellar mass value adopted for the yield computation is: $M_{500}^{\mathrm{star}} = 2.86 \pm 0.44 + [+0.00,-1.14] \times 10^{12}$ M$_{\odot}$.

\subsection*{A2589}
A stellar mass measurement for this cluster is also provided by \citet{zhang11_corr}, who report $M_{500}^{\mathrm{star}} = (5.12 \pm 0.56) \times 10^{12}$ M$_{\odot}$, based on \citet{salpeter55} as reference IMF and assuming a total mass of $M_{500} \simeq 3.03 \times 10^{14}$ M$_{\odot}$. As done for MKW3s, we adjusted this value by dividing it by 2 to account for the difference in IMF, and then multiplied it by 0.85 to rescale it to our adopted $M_{500}$ value. Statistical uncertainties associated with the total masses were also propagated in the conversion to derive the final error.

We also defined lower and upper bounds for the stellar mass of this cluster. To obtain the lower limit, we divided the reference value by $1.3$, following the calibration between \citet{zhang11_corr} and \citet{ghizzardi21}. The resulting value was then multiplied by $1.6$, accounting for the offset between \citet{ghizzardi21} and \citet{kravtsov18}, to derive the upper limit. Therefore, accounting for both statistical and systematic uncertainties, the final stellar mass value adopted for the yield computation is: $M_{500}^{\mathrm{star}} = 2.20 \pm 0.63 + [+0.51,-0.51] \times 10^{12}$ M$_{\odot}$. 

\subsection*{Hydra A}
\begin{figure}
            \centering
            \includegraphics[width = 0.456\textwidth]{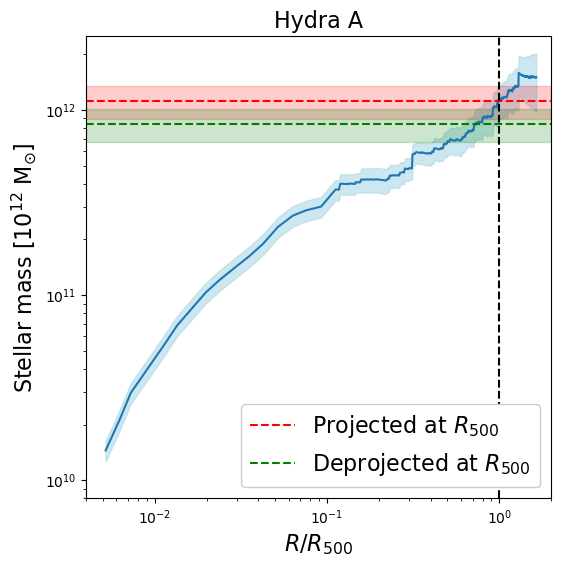}
            \caption{\footnotesize Cumulative projected stellar mass profile of Hydra A, produced within the X-COP framework following the methodology described in \citet{vanderburg15}. The red dashed line and shaded area indicate the best-fit value and associated uncertainty of the projected stellar mass at $R_{500}$. The green line and shaded region represent the corresponding deprojected stellar mass within the same overdensity.}
            \label{hydra_sm}
\end{figure}

For this cluster, we relied on a stellar mass profile produced as part of the X-COP project \citep{ghizzardi21}, following the methodology described in \citet{vanderburg15}. The cumulative stellar mass profile had not been published previously and is now shown in blue in Fig.~\ref{hydra_sm}. The profile was already calculated assuming \citet{chabrier03} as reference IMF. We interpolated the profile at our best-fit $R_{500}$ value and subsequently applied a correction factor of 0.75 to convert the projected stellar mass to the corresponding deprojected value within the same overdensity, as detailed in \citet{ghizzardi21}.

Starting from the deprojected stellar mass measured within $R_{500}$, we applied the cross-calibration between \citet{ghizzardi21} and \citet{kravtsov18} by multiplying the value by 1.6, in order to estimate an upper limit and account for potential systematic uncertainties. No calibration towards lower stellar mass values could be defined in this case. The final stellar mass adopted for Hydra A, including both statistical and systematic uncertainties, is: $M_{500}^{\mathrm{star}} = 0.84 \pm 0.17 + [+0.22,-0.00] \times 10^{12} \, \mathrm{M}_{\odot}$.

\end{appendix}

\end{document}